\documentclass[twocolumn]{autart}    

\usepackage{xcolor}
\usepackage{cite}
\usepackage{amsmath,amssymb,amsfonts}
\usepackage{algorithmic}
\usepackage{graphicx}
\usepackage{textcomp}
\usepackage{xcolor}
\usepackage{tikz}
\usepackage{subfig}
\usepackage{graphicx}    
\def\BibTeX{{\rm B\kern-.05em{\sc i\kern-.025em b}\kern-.08em
    T\kern-.1667em\lower.7ex\hbox{E}\kern-.125emX}}

\newcommand{\m}[1]{\mathbf{#1}}
\newcommand{\mc}[1]{\mathcal{#1}}
\newcommand{\mb}[1]{\mathbb{#1}}

\newcommand{\bm}[1]{\boldsymbol{#1}}

\newtheorem{remark}{Remark}
\newtheorem{lemma}{Lemma}
\newtheorem{theorem}{Theorem}
\newtheorem{corollary}{Corollary}

\begin{document}

\begin{frontmatter}

\title{Bearing-Only Solution for Fermat-Weber Location Problem: Generalized Algorithms} 

\thanks[footnoteinfo]{A part of this paper was presented at \cite{LPNM2023FWLP}. Corresponding author: Minh Hoang Trinh.}

\author[LPNM]{Nhat-Minh Le-Phan}\ead{minh.lephannhat@hust.edu.vn},    
\author[LPNM]{Phuoc Doan Nguyen}\ead{phuoc.nguyendoan@hust.edu.vn},  
\author[GIST]{Hyo-Sung Ahn}\ead{hyosung@gist.ac.kr},  
\author[THM]{Minh Hoang Trinh}\ead{minhtrinh@ieee.org}             
\address[LPNM]{Department of Automation Engineering, School of Electrical and Electronic Engineering, Hanoi University of Science and Technology, Hanoi, Vietnam.}  
\address[GIST]{School of Mechanical Engineering, Gwangju Institute of Science and Technology (GIST), Gwangju, Republic of Korea.}  
\address[THM]{AI Department, FPT University, Quy Nhon AI Campus, Binh Dinh, Vietnam}


\begin{abstract}  

This paper presents novel algorithms for the Fermat-Weber Location Problem, guiding an autonomous agent to the point that minimizes the weighted sum of Euclidean distances to some beacons using only bearing measurements. The existing results address only the simple scenario where the beacons are stationary and the agent is modeled by a single integrator. In this paper, we propose a number of bearing-only algorithms that let the agent, which can be modeled as either a single-integrator or a double-integrator, follow the Fermat-Weber point of a group of stationary or moving beacons. The theoretical results are rigorously proven using Lyapunov theory and supported with simulation examples.
\end{abstract}

\end{frontmatter}

\section{Introduction}
The Fermat-Weber Location Problem (FWLP) is a classical problem in operational research, focusing on identifying a position called Fermat-Weber point that minimizes the weighted sum of distances to a set of beacons. Traditionally, this problem has been studied with the assumption that the beacons are stationary and not collinear \cite{Plastria2011}.
In 1937, Hungarian mathematician Endre Weiszfeld proposed a numerical algorithm for finding the minimum, known as the Fermat-Weber point, which has since been widely studied by other researchers \cite{Kuhn1973,Brimberg1995,Plastria2011}. However, Weiszfeld's original analysis did not address the weighted problem, assuming instead that all weights were equal to one. The corresponding results for the weighted case, with nearly identical analysis, can be found in \cite{Beck2015JOTA}. Additionally, although Weiszfeld assumed that the initial estimator was distinct from any beacon location, it is possible for the estimator to become collocated with a beacon at a specific moment and get stuck there \cite{Kulin1962}.  

This paper considers the FWLP in the context of the guidance problem, focusing on scenarios where a robot or an autonomous agent navigates to the Fermat-Weber point of beacons for surveillance purposes, despite GPS information being unavailable or unreliable. Thus, the agent can only sense relative information from the beacons. Although there are many complex active sensors available, low-cost and lightweight passive onboard sensors are commonly used in small-sized agents to simplify the required hardware systems. Nevertheless, in many guidance problems, bearing measurements are one of the few types of information available from passive sensors. Hence, we focus specifically on the case where sensors can only obtain bearing measurements. Compared to distance and other types of measurements, bearing sensing capability is a minimal requirement for the agent. In the real world, bearing measurements can be obtained by an on-board camera. Bearing-only algorithms have been extensively researched for various navigation, guidance, and control problems. In \cite{TRINH2016CEP}, a bearing-only measurement-based guidance law was introduced for the agent to reach its desired location, characterized by a set of feasible bearing angles to three stationary beacons in the plane. 
In \cite{Li2023TRO}, bearing-only guidance laws was proposed for the problem of target following. The bearing-only observers are designed to estimate the motion of the target, which is also studied in \cite{parayil2018JGCD,SABET2016Ocean}. A wall-following UAV guidance law was proposed in \cite{suresh2020JGCD}, where bearing information, also referred to as line-of-sight (LOS), was utilized for a proportional navigation law. It is notable that in several algorithms, the bearing rate is required. For example, in a missile guidance problem, the line-of-sight (LOS) rate can be obtained from the seeker and plays a vital role in every guidance law \cite{zarchan2012tactical}. The bearing-only formation tracking control problem was considered in \cite{Zhao2019tac,Zhao2015tac,lin2023IJRN}. In these papers, the objective is to design a distributed bearing-only control law for multiple agents, allowing them to eventually achieve a desired formation. The desired formation is defined by a set of desired bearing constraints among the agents and the leaders' positions. The main difference between formation tracking and the FWLP is that, the Fermat-Weber point is implicitly defined by an algebraic equation involving the beacons and cannot be determined from the beacons' positions and the bearing measurements. The agent needs to use only bearing information to eventually detect and track the point which satisfies the algebraic equation.

Bearing-only solution to the Fermat-Weber problem was first proposed in \cite{Trinh2015Fermat}, where a single-integrator agent model was considered. This simple, continuous-time solution can be considered as an application of the steepest gradient descent algorithm. In \cite{LPNM2023FWLP}, the authors utilized the dimensional lifting technique that rules out the possibility of collision between the agents and the beacons while asymptotically stabilizing the Fermat-Weber point. Furthermore, the finite time control law and the analysis of robustness to small imprecise measurements in two-dimensional space were also proposed. On the other hand, in many problems, such as surveillance, where the agent needs to move to a position that optimizes supervision of all beacons, it is natural to consider that the beacons, which may represent a formation of ground mobile robots, are in motion. Thus, it is necessary to study the Fermat-Weber problem with moving beacons. Another issue is that the algorithms proposed in \cite{Trinh2015Fermat,LPNM2023FWLP} are only applicable to the single-integrator agent model. For many practical systems, a second-order dynamics provides a more appropriate model. 

The main contribution of this article is to propose novel bearing-only algorithms for the FWLP under different conditions. First, we extend existing work for the single-integrator agent model to include global exponential stability, robustness analysis against measurement errors in an arbitrary dimensional space. We also provide a proof of finite-time convergence, which was not presented in \cite{LPNM2023FWLP}. Furthermore, we propose algorithms when the beacons are in motion, thus, greatly extend the applicability of the problem in comparison with the existing works. Second, the algorithms for the double-integrator agent model are designed to accommodate both stationary and moving beacons, under the assumption that the agent can also access to its own velocity. The stability analysis for double-integrators are established based on Barbalat's lemma. The proposed algorithms in this paper represent a crucial step in extending the application of FWLP to practical tasks such as surveillance, aerial-ground cooperative control, and autonomous guidance of a vehicle. 

The remainder of this paper is organized as follows: Section \ref{sec:Notation} presents notations and formulates the problem. Sections \ref{sec:Single-Integrator} and \ref{sec: Double-Integrator} propose the control laws that solve the FWLP for the single- and double-integrator agent models, respectively. Simulation results are  provided in Section \ref{sec:simulation}. Finally, Section \ref{sec:conclusion} concludes our article.





\section{Notations and Problem Statement}\label{sec:Notation}
\subsection{Notations}
The set of real numbers is denoted by $\mb{R}$. For a vector $\m{x}=[x_1, \ldots, x_d]^\top\in\mb{R}^d$, $\|\m{x}\|$ denotes its Euclidean norm, and $|\m{x}| = [|x_1|, \ldots, |x_d|]^\top$. $\m{1}_n \in \mb{R}^n$ is the vector of all ones, $\m{I}_d$ denotes a $d\times d$ identity matrix, and $\m{0}$ denotes a zero vector/matrix with proper size. A matrix $\m{A}$ has the kernel and image denoted by ker$(\m{A})$ and im$(\m{A})$, respectively. For a vector $\m{x} \neq \m{0}$, we can define a corresponding projection matrix $\m{P}_{\m{x}} = \m{I}_d - \frac{\m{x}}{\|\m{x}\|}\frac{\m{x}^\top}{\|\m{x}\|}$. Then, $\m{P}_{\m{x}}= \m{P}_{\m{x}}^\top \geq 0$. Moreover, $\text{ker}(\m{P}_{\m{x}})=\text{im}(\m{x})$. In addition, $\lambda_{\min}(\cdot)$ and $\lambda_{\max}(\cdot)$ denote the smallest and the largest eigenvalues of the corresponding matrix, respectively.

Let us consider an autonomous agent, initially located at $\m{p}(0)\in \mb{R}^{d}$. Let the positions of the beacons be correspondingly given as $\m{p}_i \in \mb{R}^{d}$. Denote the displacements, distances, and bearing vectors with regard to the $i$-th beacon $i=1, \ldots, n,$ when the agent is at $\m{p}(t)$ as
\begin{subequations}
\begin{align}
\m{z}_i &= \m{p}_i - \m{p},\\
d_i &= \|\m{p}_i - \m{p}\|,\\
\m{g}_{i} &= \frac{\m{p}_i-\m{p}}{\|\m{p}_i-\m{p}\|} = \frac{\m{z}_i}{\|\m{z}_i\|}.
\end{align}
\end{subequations}

When the agent collides with beacon $i$, i.e., $\m{p} = \m{p}_i$, we have a convention that $\m{g}_i = \m{0}_d$. It is clear that $\mathbf{P}_{\m{g}_i}$ is symmetric. Furthermore, $\mathbf{P}_{\m{g}_i}^2=\mathbf{P}_{\m{g}_i} \geq 0$ (idempotent and positive semidefinite), $\text{ker}(\mathbf{P}_{\m{g}_i}) = \text{span}(\m{g}_i)$ and
$\mathbf{P}_{\m{g}_i}$ has one zero eigenvalue and $d-1$ unity eigenvalues \cite{Zhao2015tac}. 

\subsection{Problem Statement}
We consider two types of dynamics of agent
\begin{itemize}
    \item Single-integrator model
\begin{equation}\label{eq:single}
    \Dot{\m{p}}(t) = \m{u}(t),
\end{equation}
    \item  Double-integrator model
\begin{equation}\label{eq:double}
\begin{aligned}
    \Dot{\m{p}}(t) &= \m{v}(t),\\
    \Dot{\m{v}}(t) &= \m{u}(t),
\end{aligned}
\end{equation}
\end{itemize}
where $\m{v}(t) \in \mb{R}^d$ is the velocity and $\m{u}(t) \in \mb{R}^d$ is the control input. 

Consider a set of $n\geq 3$ beacons that may be stationary or moving. We assume that the beacon's positions are not collinear. In this article, our objective is to design bearing-only control laws\footnote{In this paper, bearing-only control laws mean algorithms that only require the bearing measurement between the agent and beacons rather than other relative information.} $\m{u}(t)$ for each type of agent model so that the agent will eventually track the point $\m{p}^* \in \mb{R}^d$ that solves the Fermat-Weber problem, i.e., minimizes the following function:
\begin{equation}
f(\m{p})=\sum_{i=1}^n \omega_i \|\m{p}-\m{p}_i\|,
\end{equation}
where $\omega_i>0$ are positive constants. To this end, the following lemma is well-known

\begin{lemma}\cite{Plastria2011} \label{lem:existence}
There exists a unique Fermat-Weber point $\m{p}^*$ that minimizes $f(\m{p})$ if the following inequalities
\begin{equation} \label{eq:equib_pt2}
\left|\left| \sum \limits_{i = 1; i \neq k}^n \omega_i \m{g}_i  \right|\right| > \omega_k,~k=1,\ldots,N,
\end{equation}
hold, for all $\m{p} = \m{p}_k$. The Fermat-Weber point $\m{p}^*$ is different from all $\m{p}_i$, $i = 1,\ldots, n$, and satisfies
\begin{equation} \label{eq:equib_pt1}
\sum \limits_{i = 1}^n \omega_i \m{g}_i^* =\sum \limits_{i = 1}^n \omega_i \frac{\m{p}_i-\m{p}^*}{\|\m{p}_i-\m{p}^*\|} = \m{0}_d.
\end{equation}
\end{lemma}

Equality (\ref{eq:equib_pt1}) can be rewritten as 
$$\mathbf{p}^* = \left( \sum \limits_{i = 1}^n \frac{\omega_i}{d_i^*} \right)^{-1}\sum \limits_{i = 1}^n \frac{\omega_i}{d_i^*}\mathbf{p}_i,$$
where $d_i^* = \|\mathbf{p}_i - \mathbf{p}^*\| > 0$. Thus, $\mathbf{p}^*$ always lies in the convex hull of the beacons. 

\section{Single-Integrator Agent}\label{sec:Single-Integrator}

This section consider the single-integrator agent model \eqref{eq:single}. To begin with, we revisited the bearing-only algorithm introduced in \cite{Trinh2015Fermat}, which assumes stationary beacons, and proposed several new results. Next, we design the control laws in the case of moving beacons, where their velocity is assumed to be unknown.

\subsection{Stationary Beacons: The Gradient Control Law Revisited and Novel Results} 

For simplicity, we assumed that the Fermat-Weber point is not a beacon, i.e., $\m{p}^* \neq \m{p}_i \ \ \forall i =1,\dots,n$. In such a case, \cite{Trinh2015Fermat} proposes the following bearing-only control law

\begin{align} \label{eq:control_law_gradient_descent}
\m{u} = \sum_{i=1}^n \omega_i \frac{{\m{p}}_i-\m{p}}{\|{\m{p}}_i-\m{p}\|} =\sum_{i=1}^n \omega_i {\m{g}}_i.
\end{align}

Indeed, due to the fact that $\frac{\partial {d}_i}{\partial \m{p}} = -{\m{g}}_i^\top,~\frac{\partial {\m{g}}_i}{\partial \m{p}} = - \frac{\m{P}_{{\m{g}}_{i}}}{{d}_i},$ the control law \eqref{eq:control_law_gradient_descent} can be rewritten as
\begin{equation} \label{eq:gradient-control-law}
\m{u} = -\nabla_{\m{p}} {f}(\m{p}) = -\sum_{i=1}^n \omega_i \left(\frac{\partial {d}_i}{\partial \m{p}}\right)^\top = \sum_{i=1}^n \omega_i {\m{g}}_{i},
\end{equation}
which is the steepest gradient descent corresponding to the objective function
${f}(\m{p}) =  \sum_{i=1}^n \omega_i {d}_i$. 
In the case of stationary beacons and single-integrator model agents, it was rigorously proved that (\ref{eq:equib_pt1}) can solve the FWLP.

\subsubsection{Exponential Stability}

First, by denoting $\bm{\delta}(t) = \m{p}-\m{p}^*$,  $\m{z}_i^* = \m{p}_i - \m{p}^*$, $\m{z}=[\m{z}_1^\top,\dots,\m{z}_2^\top]^\top$, $\m{z}^*=[\m{z}_1^{*\top},\dots,\m{z}_2^{*\top}]^\top$, ${\mathbf{g}} = [{\mathbf{g}}_1^\top,\dots, {\mathbf{g}}_n^\top]^\top$, $\mathbf{g}^* = [{\mathbf{g}}_1^{*\top},\dots,{\mathbf{g}}_n^{*\top}]^\top$ and $\bar{\m{W}} = \text{blkdiag}(\omega_i\mathbf{I}_{d})$, we prove the following lemma.
\begin{lemma}\label{lemma 3}
The following holds
\begin{subequations}\label{eq:bound}
\begin{align} 
\m{z}^\top\bar{\m{W}}({\m{g}}-{\m{g}}^*) &\geq 0\\
\m{z}^{*\top}\bar{\m{W}}({\m{g}}-{\m{g}}^*) &\leq 0,
\end{align}
\end{subequations}
where the equal signs occur when $\m{p} = \m{p}^*$.
\end{lemma}
\begin{pf} It can be obtained that 
\begin{subequations}
\begin{align}
\m{z}^\top\bar{\m{W}}({\m{g}}-{\m{g}}^*) &=  \sum_{i=1}^n \omega_i {d}_i(1-{\mathbf{g}}_i^\top{\mathbf{g}}_i^*)\label{eq: 12a} \\
\m{z}^{*\top}\bar{\m{W}}({\m{g}}-{\m{g}}^*) &=  \sum_{i=1}^n \omega_i {d}_i^* ({\mathbf{g}}_i^{*\top}{\mathbf{g}}_i-1).\label{eq: 12b}
\end{align}
\end{subequations}
As $-1 \leq\mathbf{g}_i^{*\top}{\mathbf{g}}_i \leq 1$, the lemma is completely proven. 
\hfill$\blacksquare$
\end{pf}

\begin{lemma}\label{lemma: boundness_di}
The distance $d_i$ is constrained by the following lower and upper bounds 
\begin{subequations}
\begin{align}
d_i &\leq \|\bm{\delta}(t)\|+\max_{i\in \{1,\ldots,n\}} {d}_i^*,\label{eq: 13a}\\
d_i & \geq \|\bm{\delta}(t)\| - \max_{i\in \{1,\ldots,n\}} {d}_i^*,\label{eq: 13b}\\
d_i & \geq \min_{i\in \{1,\ldots,n\}} {d}_i^*-\|\bm{\delta}(t)\|.\label{eq: 13c}
\end{align}
\end{subequations}
\end{lemma}
\begin{pf} 
 Note that $\m{p}-\m{p}_i = (\m{p}-\m{p}^*) - (\m{p}_i-\m{p}^*)=\bm{\delta}(t)-\m{z}_i^*$. Therefore, this lemma can be derived by using the triangle inequality.
\hfill$\blacksquare$
\end{pf}
\begin{corollary}\label{cor:1a}
The following holds
\begin{equation}
\begin{aligned}
\m{z}^\top\bar{\m{W}}({\m{g}}-{\m{g}}^*) &\geq \frac{\lambda_{\min}(\sum_{i = 1}^n \omega_i\mathbf{P}_{{g}^*_i})\|\bm{\delta}(t)\|^2}{2\big(\|\bm{\delta}(t)\|+\max_{i\in \{1,\ldots,n\}} {d}_i^*\big)}.
\end{aligned}
\end{equation}
\end{corollary}
\begin{pf}
Note that $\mathbf{P}_{{g}^*_i}\m{z}_i^* = 0$ for all $i\in \{1,\dots,n\}$. One has
\begin{align}
    \bm{\delta}^\top\left(\sum_{i = 1}^n \omega_i\mathbf{P}_{{g}^*_i}\right)\bm{\delta} &= \sum_{j = 1}^n \omega_i(\m{z}_i^*-\m{z}_i)^\top\mathbf{P}_{{g}^*_i}(\m{z}_i^*-\m{z}_i) \nonumber\\
    &= \sum_{j = 1}^n \omega_i\m{z}_i^\top\mathbf{P}_{{g}^*_i}\m{z}_i \nonumber\\
    &= \sum_{i=1}^n\omega_i {\m{z}}_i^\top(\m{I}_{d+1}-{\m{g}}_i^{*}{\m{g}}_i^{*\top}){\m{z}}_i \nonumber\\
    &= \sum_{i=1}^n\omega_i {d}_i^2(1-({\m{g}}_i^{\top}{\m{g}}_i^{*})^2) \nonumber\\
    &=\sum_{i=1}^n\omega_i {d}_i^2(1-{\m{g}}_i^{\top}{\m{g}}_i^{*})(1+{\m{g}}_i^{\top}{\m{g}}_i^{*}) \nonumber\\
    &\leq 2\max_{i\in \{1,\ldots,n\}} {d}_i\sum_{i=1}^n\omega_i {d}_i(1-{\m{g}}_i^{\top}{\m{g}}_i^{*}) \nonumber\\ &=2\max_{i\in \{1,\ldots,n\}} {d}_i{\m{z}}^\top\bar{\m{W}}({\m{g}}-{\m{g}}^*). \label{eq:co1a}
\end{align}
The matrix $\sum_{i = 1}^n \omega_i\mathbf{P}_{{g}^*_i}$ is symmetric and positive definite as beacons are not colinear. Thus, 
$\lambda_{\min}(\sum_{i = 1}^n \omega_i\mathbf{P}_{{g}^*_i})\|\bm{\delta}(t)\|^2 \leq \bm{\delta}^\top(\sum_{i = 1}^n \omega_i\mathbf{P}_{{g}^*_i})\bm{\delta}$ where $\lambda_{\min}(\sum_{i = 1}^n \omega_i\mathbf{P}_{{g}^*_i}) > 0$. Combining with (\ref{eq: 13a}), the corollary is completely proven. \hfill$\blacksquare$
\end{pf}

\begin{theorem}
Suppose that there is no collision among the agent and beacons. Under the control law \eqref{eq:control_law_gradient_descent}, for all $\m{p}(0)=[p_{1}(0),\ldots,p_{d}(0)]^\top$, $\m{p}(t)$ exponentially asymptotically converges to ${\m{p}}^*$.    
\end{theorem}
\begin{pf} Firstly, we prove the global asymptotic stability of $\m{p}^*$. Consider the Lyapunov function
$$V = \frac{1}{2}\bm{\delta}^\top\bm{\delta},$$
which is positive definite, continuously differentiable almost everywhere, and radially unbounded. By taking the derivative of $V$, the following yields
\begin{equation}
\begin{aligned}
    \dot{V} &= \bm{\delta}^\top(t)\Dot{\m{p}}(t) = \sum_{i=1}^n \omega_i(\m{p}(t)-\m{p}^*)^\top {\m{g}}_i\\
    &= \sum_{i=1}^n \omega_i(\m{p}(t)-\m{p}_i+\m{p}_i-\m{p}^*)^\top {\m{g}}_i\\
    &=-{\m{z}}^\top\bar{\m{W}}\big({\mathbf{g}}-{\mathbf{g}}^*\big)+{\m{z}}^{*\top}\bar{\m{W}}\big({\mathbf{g}}-{\mathbf{g}}^*\big)
\end{aligned}
\end{equation}
It is clearly seen that
$\dot{V} \leq  0$. Clearly, $\dot{V} = 0$ if and only if ${\m{g}}_i = \m{g}_i^*,\forall i = 1,\ldots, n$, i.e., $\m{p} = {\m{p}}^*$, thereby demonstrating the negative definiteness of $\dot{V}$. As a result, $\|\bm{\delta}(t)\| \leq \|\bm{\delta}(0)\|$ for all $t > 0$.

Secondly, by applying Lemma \ref{lemma 3} and Corollary \ref{cor:1a}, we have
\begin{equation}
\begin{aligned}
    \Dot{V} &\leq -{\m{z}}^\top\bar{\m{W}}\big({\mathbf{g}}-{\mathbf{g}}^*\big)\\
    &\leq -\frac{\lambda_{\min}(\sum_{i = 1}^n \omega_i\mathbf{P}_{{g}^*_i})\|\bm{\delta}(t)\|^2}{2\big(\|\bm{\delta}(t)\|+\max_{i\in \{1,\ldots,n\}} {d}_i^*\big)}\\
    &\leq -\underbrace{\frac{\lambda_{\min}(\sum_{i = 1}^n \omega_i\mathbf{P}_{{g}^*_i})}{\big(\|\bm{\delta}(0)\|+\max_{i\in \{1,\ldots,n\}} {d}_i^*\big)}}_{\sigma > 0}\frac{\|\bm{\delta}(t)\|^2}{2}\\
    &= -\sigma V.
\end{aligned}
\end{equation}
Hence, $\m{p}^*$ is globally exponentially stable.
\hfill$\blacksquare$
\end{pf}

\begin{remark}\label{remark:collision_first_stationary}
Let $\mc{B}_R=\{ \|\m{p}-{\m{p}}^*\|<R\}$, where $R = \min_{i\in \{1,\ldots,n\}} {d}_i^* - \epsilon$, for a small $\epsilon \in (0, \min_{i\in \{1,\ldots,n\}} {d}_i^*)$. Since $\|\bm{\delta}(t)\| < \|\bm{\delta}(0)\|$ for $t > 0$, if $\m{p}(0) \in \mc{B}_R$, one has $d_i \geq \min_{i\in \{1,\ldots,n\}} {d}_i^*-\|\bm{\delta}(t)\|
\geq \min_{i\in \{1,\ldots,n\}} {d}_i^*-\|\bm{\delta}(0)\|
> \epsilon.$ Thus, the collision between the agent and the beacons can be completely avoided.
\end{remark}

{\subsubsection{Robustness against bearing error}\label{sec: robustness}}
For robustness analysis, suppose that due to measurement errors, the obtained bearing vector ${\m{g}}_i$ is deviated by a rotation matrix $\m{R}_i$, i.e., the measured bearing vectors are given as $\m{R}_i(t){\m{g}}_i$, $\forall i=1,\ldots,n$. Thus, the bearing-only control law can be written as
\begin{align} \label{eq:control_noised}
    \m{u} = \sum_{i=1}^n \omega_i \m{R}_i(t) {\m{g}}_i.
\end{align}
The following lemma was proven in \cite{Trinh2015Fermat,Plastria2011}.

\begin{lemma} \label{lem:hessian}
Let $n\geq 3$ and suppose that beacons are not collinear.  Then, 
\begin{itemize}
    \item the matrix $\m{H}(\m{p})= \frac{\partial {f}^2(\m{p})}{\partial \m{p}^2}= \sum_{i=1}^n\omega_i \frac{\m{P}_{{\m{g}}_{i}}}{{d}_i}$ is positive definite for all $\m{p}\in \mc{B}_R$.
    \item ${f}(\m{p})$ is strongly convex on $\mc{B}_R$, i.e., $\forall \m{p}, \m{q} \in \mc{B}_R$, 
    \begin{align} \label{eq:strong_convexity}
        {f}(\m{p})-{f}(\m{q})\geq \frac{\partial {f}(\m{p})}{\partial \m{p}} (\m{p}-\m{q}) + \frac{m}{2}\|\m{p}-\m{q}\|^2,
    \end{align}
    where $m=\min_{\m{p}\in \mc{B}_R} \lambda_1(\m{H}(\m{p}))>0$ denotes the smallest positive eigenvalue of $\m{H}$.
\end{itemize}
\end{lemma}
The robustness of the control law under measuring error is considered in the following theorem.
\begin{theorem}
Assume that $\m{p}(0) \in \mc{B}_R$ and $\m{R}_i(t)$ are positive definite with $\inf_{t\geq 0}\lambda_{\min}(\m{R}_i(t)+\m{R}_i^\top(t)) = 2\Lambda_i>0,~\forall i = 1, \ldots, n$. Under the control law \eqref{eq:control_noised}, ${\bm{\delta}}(t)=\m{p}(t)-{\m{p}}^*$ is uniformly ultimately bounded.
\end{theorem}
\begin{pf}
Consider the Lyapunov function $V=\frac{1}{2}\|\boldsymbol{\bm{\delta}}\|^2$, which is positive definite, continuously differentiable in $\mc{B}_R$, and $\alpha_1 \|\boldsymbol{\bm{\delta}}\|^2 \leq V \leq \alpha_2 \|\boldsymbol{\bm{\delta}}\|^2$, for $\alpha_1<0.5<\alpha_2$. We have
\begin{align*}
    \dot{V} &= \boldsymbol{\bm{\delta}}^\top \sum_{i=1}^n \omega_i \m{R}_i {\m{g}}_i = (\m{p}-{\m{p}}_i + {\m{p}}_i-\m{p}^*)^\top \sum_{i=1}^n \omega_i \m{R}_i {\m{g}}_i\\
    & = \sum_{i=1}^n \omega_i (-{d}_i {\m{g}}_i + {d}_i^*\m{g}_i^*)^\top \m{R}_i {\m{g}}_i\\
    &= -\sum_{i=1}^n \omega_i \Lambda_i {d}_i  + \sum_{i=1}^n \omega_i {d}_i^*(\m{g}_i^*)^\top \m{R}_i(t) \m{g}_i\\
    &\leq - \left(\min_{i\in \{1,\ldots,n\}}\Lambda_i\right) {f}(\m{p}) + {f}(\m{p}^*).
\end{align*}
Denoting $\Lambda = \min_{i\in \{1,\ldots,n\}}\Lambda_i$, 
and using the strong convexity of the function ${f}({\m{p}})$, and $\frac{\partial {f}({\m{p}}^*)}{\partial \m{p}}=\m{0}$, we can write 
\begin{align*}
    \dot{V} &\leq -\Lambda ({f}(\m{p})-{f}({\m{p}}^*)) + (1-\Lambda) {f}({\m{p}}^*)\\
    & \leq - \Lambda {\frac{\partial {f}({\m{p}}^*)}{\partial \m{p}}} \boldsymbol{\bm{\delta}}- \frac{m\Lambda}{2}\|\boldsymbol{\bm{\delta}}\|^2 + (1-\Lambda) {f}({\m{p}}^*)\\
    &= -m\Lambda V + (1-\Lambda) {f}({\m{p}}^*)\\
    &=-\mu m\Lambda V -(1-\mu) m\Lambda V + (1-\Lambda) {f}({\m{p}}^*),
\end{align*}
where $\mu \in (0,1).$ This implies that $V(t) = \frac{\|\bm{\delta}(t)\|^2}{2}$ will exponentially decrease until $V(t)  \leq \frac{(1-\Lambda) {f}({\m{p}}^*)}{(1-\mu) m\Lambda}$, i.e., $\|\bm{\delta}(t)\| \leq \sqrt{\frac{2(1-\Lambda) {f}({\m{p}}^*)}{(1-\mu) m\Lambda}}$ within a finite time. Finally, the ultimate bound of $\bm{\delta(t)}$ decreases as the measuring error $\theta_i \rightarrow 0$, i.e., as $\Lambda \rightarrow 1$. \hfill$\blacksquare$
\end{pf}

\begin{remark}
Consider $d =2$, and the bearing vector $\m{g}_i(t)$ is deviated by an angle $\theta_i(t)$. The assumption of the positive definiteness of $\mathbf{R}_i(t) =
\begin{bmatrix}
    &\text{cos}(\theta_i(t)) 
    &&-\text{sin}(\theta_i(t))\\
    &\text{sin}(\theta_i(t)) 
    &&\text{cos}(\theta_i(t))\\
\end{bmatrix}
$ implies that, despite measurement errors, any measured bearing vector should not be deflected by an angle $|\theta_i(t)| \geq \frac{\pi}{2}$ from the corresponding actual bearing vector. 
\end{remark}

\subsubsection{Finite Time Control Law}
The finite time control law is proposed as follows
\begin{align} \label{eq:finite_time}
\m{u} = \text{sig}^{a}\left(\sum_{i=1}^n \omega_i {\m{g}}_i \right),
\end{align}
where $\text{sig}^a(x) = \text{sgn}(x)|x|^a$, and $a \in (0,1)$. Before considering the stability under the proposed control law, we have the following lemma

\begin{lemma}\label{lemma: closed region}
Every point $\mathbf{p}(t)\in \mc{B}_R$ satisfies
$$\mathbf{g}_i^\top{\mathbf{g}}_i^* > 0, \ \ \forall i = 1,\ldots, n.$$
\end{lemma}
\begin{pf} If $\mathbf{p}(t)\in \mc{B}_R$, we have
$\|\m{p}(t)-\m{p}^*\|<\|\m{p}_i - \m{p}^*\| \ \ \forall i = 1,\ldots, n$,
which is equivalent to
\begin{equation*}
\begin{aligned}
(\m{p}-\m{p}^*)^\top(\m{p}-\m{p}^*) &< (\m{p}_i-\m{p}+\m{p}-\m{p}^*)^\top(\m{p}_i-\m{p}^*)\\
(\m{p}-\m{p}^*)^\top(\m{p}-\m{p}_i) &<(\m{p}_i-\m{p})^\top (\m{p}_i-\m{p}^*) \\
\bm{\delta}^\top\m{z}_i &< \m{z}_i^\top\m{z}_i^*
\end{aligned}
\end{equation*}
If $\bm{\delta}^\top\m{z}_i \geq 0$, we have $\m{z}_i^\top\m{z}_i^*>0$. Suppose that there exists beacon $i$ so that $(\m{p}-\m{p}^*)^\top(\m{p}-\m{p}_i) < 0$, consider the sum
\begin{equation*}
\begin{aligned}
\bm{\delta}^\top\m{z}_i + \m{z}_i^\top\m{z}_i^* = \|\m{p}-\m{p}_i\|^2 > 0.  
\end{aligned}
\end{equation*}
Thus, $\m{z}_i^\top\m{z}_i^* = d_id_i^*{\mathbf{g}}_i^\top{\mathbf{g}}_i^* >0, \forall i$. From the definition of $\mc{B}_R$, $d_i,d_i^* > 0 \ \forall i$, and the proof follows.
\hfill$\blacksquare$
\end{pf}

\begin{figure}[hbt!]
\centering
\includegraphics[width=0.45\textwidth]{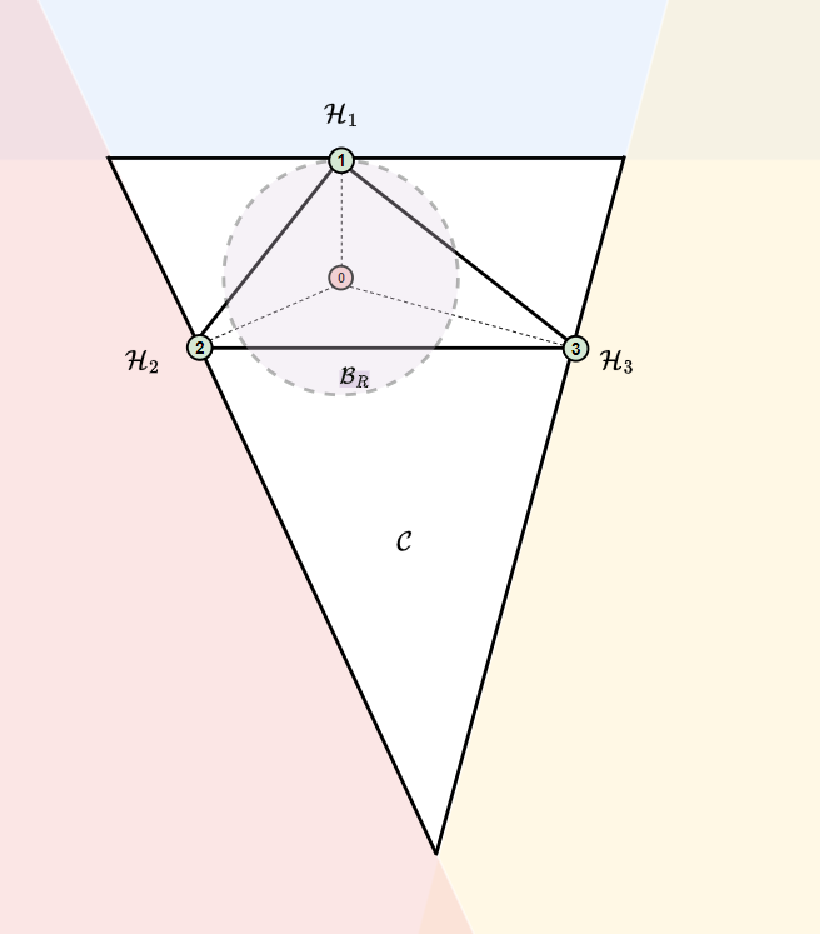}
\caption{An example of the sets $\mc{C}$ and $\mc{B}_R$ of 3 beacons in 2D space, where the Fermat-Weber point is indexed by 0.}\label{fig:triangle}
\end{figure}

\begin{remark}
Define the set $\mathcal{C} = \{\mathbf{p}(t)\in \mathbb{R}^{d}| \ \  {\mathbf{g}}_i^\top{\mathbf{g}}_i^* \geq 0 \ \ \forall i, i = 1,\ldots, n.$ This set $\mathcal{C}$ can be represented as $\mathcal{C}= \bigcap_{i=1}^n \mathcal{H}_i$, where $\mathcal{H}_i = \{\mathbf{p} \in \mathbb{R}^{d+1}| \ \ {\mathbf{g}}_i^\top{\mathbf{g}}_i^* \geq 0\}$ is actually a half-space. Thus, $\mathcal{C}$ is a convex polytope \cite{Rockafellar1970}. For instance, Fig. \ref{fig:triangle} illustrates an example of the sets $\mc{C}$ and $\mc{B}_R$ of three beacons on the plane.
\end{remark}
 
\begin{corollary}\label{co:2}
{For every $\mathbf{p}(t) \in \mc{B}_R$,} the following holds
\begin{equation}
{\m{z}}^\top\bar{\m{W}}({\m{g}}-{\m{g}}^*) \leq \frac{\lambda_{\max}(\sum_{i = 1}^n \omega_i\mathbf{P}_{{g}^*_i})\|\bm{\delta}(t)\|^2}{\min_{i\in \{1,\ldots,n\}}{d}_i(t)}
\end{equation}
\end{corollary}
\begin{pf} From Corollary \ref{cor:1a} and Lemma \ref{lemma: closed region}, one has
\begin{align}
 \bm{\delta}^\top\left(\sum_{i = 1}^n \omega_i\mathbf{P}_{{g}^*_i}\right)\bm{\delta} &=\sum_{i=1}^n\omega_i {d}_i^2(1-{\m{g}}_i^{\top}{\m{g}}_i^{*})(1+{\m{g}}_i^{\top}{\m{g}}_i^{*}) \nonumber\\
 &\geq \min_{i\in \{1,\ldots,n\}}d_i\sum_{i=1}^n\omega_i {d}_i(1-{\m{g}}_i^{\top}{\m{g}}_i^{*}) \nonumber\\
 &=\min_{i\in \{1,\ldots,n\}}d_i{\m{z}}^\top\bar{\m{W}}({\m{g}}-{\m{g}}^*).
\end{align}
Since $\bm{\delta}^\top(\sum_{i = 1}^n \omega_i\mathbf{P}_{{g}^*_i})\bm{\delta}\leq \lambda_{\max}(\sum_{i = 1}^n \omega_i\mathbf{P}_{{g}^*_i})\|\bm{\delta}(t)\|^2$, the corollary can be immediately obtained.
\hfill$\blacksquare$
\end{pf}

\begin{theorem}\label{thm:finite time}
Suppose that there is no collision among the agent and beacons. Under the control law \eqref{eq:finite_time}, $\m{p}(t)$ globally asymptotically converges to ${\m{p}}^*$. Moreover, {if $\m{p}(0) \in \mc{B}_R$}, $\m{p}(t) \rightarrow {\m{p}}^*$ in finite time.
\end{theorem}

\begin{pf} 
Consider the Lyapunov function $V = {\m{z}}^\top\bar{\m{W}}({\m{g}}-{\m{g}}^*).$ From Lemma \ref{lemma 3}, $V \geq 0$ and $V=0$ if and only if $\m{p}={\m{p}}^*$. It can be seen that ${\m{z}}^\top\bar{\m{W}}\dot{{\m{g}}} = \sum_{i=1}^n\omega_i{\m{z}}_i\dot{{\m g}}_i= 0$. The derivative of $V$ is as follows
\begin{align}
\dot{V} &= ({\m{g}}-{\m{g}}^*)^\top\bar{\m{W}}\Dot{{\m{z}}} \nonumber\\
&=-\big(\sum\nolimits_{i=1}^n \omega_i(\m{g}_i-\m{g}_i^*)\big)^\top\dot{\m{p}}\nonumber\\
&=-\sum\nolimits_{i=1}^n \omega_i(\m{g}_i-\m{g}_i^*)^\top\text{sig}^{a}\left(\sum\nolimits_{i=1}^n \omega_i {\m{g}}_i\right) \nonumber\\
&= - \sum_{k=1}^d \left|\left[\sum\nolimits_{i=1}^n \omega_i ({\m{g}}_i - \m{g}_i^*)\right]_k \right|^{\alpha+1} \nonumber\\ 
&= - \sum\nolimits_{k=1}^d \left(\left[\sum_{i=1}^n \omega_i ({\m{g}}_i - \m{g}_i^*)\right]_k^2 \right)^{\frac{\alpha+1}{2}} \nonumber
\end{align}
Using the inequality in \cite{Wang2010finite}, we have
\begin{align*}
\dot{V} &\leq - d^{\frac{1-\alpha}{2}}\left( \sum\nolimits_{k=1}^d \left[\sum_{i=1}^n \omega_i ({\m{g}}_i - \m{g}_i^*)\right]_k^2 \right)^{\frac{\alpha+1}{2}} \nonumber\\
& \leq - d^{\frac{1-\alpha}{2}}\left( \left|\left|\sum\nolimits_{i=1}^n \omega_i ({\m{g}}_i - \m{g}_i^*)\right|\right|^2 \right)^{\frac{\alpha+1}{2}}\leq 0,
\end{align*}
Based on LaSalle invariance principle, $\m{p}(t) \to \Omega = \{\m{p} \in \mb{R}^{d}|~ \sum_{i=1}^n\omega_i{\m{g}}_i = \m{0}_{d+1}\} = \{{\m{p}}^*\}$ globally asymptotically. In order to prove the second part of this theorem, we consider the case when $\m{p}(t) \in \mc{B}_R$. 
From the Cauchy-Schwartz inequality, we have
\begin{align*}
\left|\left|\sum_{i=1}^n\omega_i ({\m{g}}_i-{\m{g}}_i^*)\right|\right|^2 &= \frac{\|\bm{\delta}\|^2\|\sum_{i=1}^n\omega_i ({\m{g}}_i-{\m{g}}_i^*)\|^2}{\|\bm{\delta}\|^2} \\
&\geq \frac{\big(\bm{\delta}^\top\sum_{i=1}^n\omega_i ({\m{g}}_i-{\m{g}}_i^*)\big)^2}{\|\bm{\delta}\|^2}\\
&= \frac{\big((\m{p}-\m{p}^*)^\top\sum_{i=1}^n\omega_i ({\m{g}}_i-{\m{g}}_i^*)\big)^2}{\|\bm{\delta}\|^2}.
\end{align*}
From the fact that 
\begin{align*}
&\big(\m{p}-\m{p}_i+\m{p}_i-\m{p}^*)^\top \sum_{i=1}^n\omega_i ({\m{g}}_i-{\m{g}}_i^*)\big)^2 \\
&= \bigg(\sum_{i=1}^n\m{z}_i^\top\omega_i ({\m{g}}_i-{\m{g}}_i^*) - \sum_{i=1}^n(\m{z}^*_i)^\top\omega_i ({\m{g}}_i-{\m{g}}_i^*)\bigg)^2 \\
&= \bigg(\m{z}^\top\bar{\m{W}}({\m{g}}-{\m{g}}^*) - \m{z}^{*\top}\bar{\m{W}}({\m{g}}-{\m{g}}^*)\bigg)^2\\ 
&\geq \big(\m{z}^\top\bar{\m{W}}({\m{g}}-{\m{g}}^*)\big)^2,
\end{align*}
one can apply (\ref{eq:co1a}) and Corollary \ref{co:2}, leading to
\begin{align*}
\left|\left|\sum_{i=1}^n\omega_i ({\m{g}}_i-{\m{g}}_i^*)\right|\right|^2 &\geq \frac{\big(\m{z}^\top\bar{\m{W}}({\m{g}}-{\m{g}}^*)\big)^2}{\|\bm{\delta}\|^2}\\
&\geq\frac{\lambda_{\min}^2(\sum_{i = 1}^n \omega_i\mathbf{P}_{{g}^*_i})\|\bm{\delta}(t)\|^2}{4\max_{i}{d}_i^2(t)} \\
&\geq \chi\frac{\min_{i}{d}_i(t)}{\max_{i}{d}_i^2(t)}\m{z}^\top\bar{\m{W}}({\m{g}}-{\m{g}}^*),
\end{align*}
where \[\chi = \frac{\lambda_{\min}^2(\sum_{i = 1}^n \omega_i\mathbf{P}_{{g}^*_i})}{4\lambda_{\max}(\sum_{i = 1}^n \omega_i\mathbf{P}_{{g}^*_i})}.\] 

As $\m{p}(t) \in \mc{B}_R$, from Lemma~\ref{lemma: boundness_di}, the lower bound of $\frac{\min_{i}{d}_i(t)}{\max_{i}{d}_i^2(t)}$ can be obtained as follows
\begin{equation*}
\begin{aligned}
     \frac{\min_{i}{d}_i(t)}{\max_{i}{d}_i^2(t)} &\geq h(\|\bm{\delta}(t)\|) = \frac{\min_{i} {d}_i^*-\|\bm{\delta}(t)\|}{(\|\bm{\delta}(t)\|+\max_{i} {d}_i^*)^2} > 0\\
    \frac{\partial h(\|\bm{\delta}(t)\|)}{\partial \|\bm{\delta}(t)\|}&= \frac{\bm{\|\delta}(t)\|-\max_{i} {d}_i^*-2\min_{i} {d}_i^*}{\big(\|\bm{\delta}(t)\|+\max_{i} {d}_i^*\big)^3} < 0.
\end{aligned}
\end{equation*}
Hence, $h(\|\bm{\delta(t)}\|)$ increases as $\|\bm{\delta(t)}\|$ decreases. As a result, $h(\|\bm{\delta}\|)> h(R)$. It follows that
\begin{equation*}
    \dot{V} \leq - d^{\frac{1-\alpha}{2}}(\chi h(R))^{\frac{\alpha+1}{2}}V^{\frac{\alpha+1}{2}}= - \kappa V^{\frac{\alpha+1}{2}}.
\end{equation*}
Based on \cite{Haddad2008finite}, as $0< \frac{\alpha+1}{2}<1$, it follows that $V(t) \to 0$ in finite time. Thus, $\m{p}(t) = \m{p}^*$ if $\m{p}(0) \in \mc{B}_{\epsilon}$, $\forall t\geq T$, where $T\le \frac{2V(0)^{\frac{1-\alpha}{2}}}{\kappa^{\frac{\alpha+1}{2}}(1-\alpha)}$. 
\hfill$\blacksquare$
\end{pf}

\subsection{Moving Beacons}
We suppose that the beacons are moving with velocity $\m{v}_i(t)=[v_1,\ldots,v_d]^\top \in \mb{R}^{d}$, i.e.,
\begin{equation}
\dot{\m{p}}_i(t) = \m{v}_i(t),~i = 1, \ldots, n,
\end{equation}
Then, the minimum $\m{p}^*$ of the function $f(\m{p})$ is also time-varying. Let the beacons move with the same known velocity, $\m{v}_i=\m{v^*},\forall i=1,\ldots, n$. Then, the relative positions between the beacons are time-invariant, and the minimum $\m{p}^*$ moves with the same velocity as the beacons. This statement can be proved by a simple change of variables ${\m{q}_i(t)} = \m{p}_i(t) - \int_{0}^{t}\m{v}(\tau)d\tau$, of which $\dot{\m{q}}_i = \m{0}_{d}$. Because the minimum $\m{q}^*$ of ${f}(\m{q})$ should satisfy all the desired vectors $\{\m{g}_i^*\}_{i=1,\ldots,n}$, one has
\begin{align}
\m{P}_{\m{g}_i^*}(\m{q}_i - \m{q}^*) = \m{0}_{d+1},~i = 1, \ldots, n.
\end{align}
This implies $\m{q}^* = \left(\sum_{i=1}^n \m{P}_{\m{g}_i^*}\right)^{-1} \sum_{i=1}^n \m{P}_{\m{g}_i^*}\m{q}_i$, and thus
\begin{align}
\dot{\m{q}}^* = \left(\sum_{i=1}^n \m{P}_{\m{g}_i^*}\right)^{-1} \sum_{i=1}^n \m{P}_{\m{g}_i^*} \dot{\m{q}}_i = \m{0}_{d+1}.
\end{align}
Hence, one has $\dot{\m{q}}^*(t) = \dot{\m{p}}^*(t) - \m{v}(t) = \m{0}_{d+1},$ or $\dot{\m{p}}^*(t) = \m{v}(t)$. Correspondingly, the Fermat-Weber point ${\m{p}}^*(t)$ also moves with velocity $\m{v}(t)$.

\subsubsection{The beacons move with the same constant velocity}
Suppose that the beacons move with unknown constant velocity, i.e., $\dot{\m{v}}^* = \m{0}_{d}$. The following adaptive control law is proposed
\begin{equation}
\label{eq:PI-control}
\begin{aligned}
\m{u}(t) & = \sum_{i=1}^n \omega_i {\m{g}}_i + \hat{\m{v}}(t),\\
\dot{\hat{\m{v}}}(t) & = k\sum_{i=1}^n \omega_i {\m{g}}_i. 
\end{aligned}
\end{equation}
where $k>0$ is a control gain and $\boldsymbol{ \hat{\m{v}}}(0)=\m{0}_{d}$. We will prove the following theorem
\begin{theorem}\label{thm:single_constant_v}
Let the beacons move with constant velocity $\m{v}^*$. Suppose that there is no collision among the agent and beacons. Under the control law \eqref{eq:PI-control}, for all $\m{p}(0)\in \mb{R}^d$, $\|\m{p}(t)-{\m{p}}^*\|\to 0$ as $t\to \infty$.
\end{theorem}
\begin{pf} Consider the function \[V = {\m{z}}^\top\bar{\m{W}}({\m{g}}-{\m{g}}^*) + \frac{\|\hat{\mathbf{v}}-\m{v}^*(t)\|^2}{2k}.\] 
According to Lemma \ref{lemma 3}, $V$ is positive definite with regard to $\bm{\delta}$ and $\hat{\m{v}}-\m{v}^*(t)$. The derivative of $V$ can be obtained as follows
\begin{align*}
\dot{V} &= ({\m{g}}-{\m{g}}^*)^\top\bar{\m{W}}\Dot{{\m{z}}} + \frac{1}{k}(\hat{\mathbf{v}}-\m{v}^*)^\top \dot{\hat{\m{v}}}\\
& = \sum_{i=1}^n \omega_i ({\m{g}}_i-{\m{g}}_i^*)^\top\left(\m{v}^* - \sum_{i=1}^n \omega_i {\m{g}}_i -\hat{\m{v}}(t) \right) \nonumber \\
&\qquad\qquad + (\hat{\mathbf{v}}-\m{v}^*)^\top \sum_{i=1}^n \omega_i {\m{g}}_i\\
& = - \left|\left|\sum_{i=1}^n \omega_i {\m{g}}_i\right|\right|^2 \leq 0.
\end{align*}
From LaSalle's invariance principle, each trajectory converges
to the largest invariance set inside $\Omega = \{\Dot{V} = 0\}$. Thus, it follows that $\sum_{i=1}^n \omega_i {\m{g}}_i \equiv \m{0}$. From Lemma~\ref{lem:existence}, the Fermat-Weber point is asymptotically tracked. Furthermore, from $\m{p}(t)\equiv{\m{p}}^*(t)$, it holds that $\m{v}(t) \equiv \m{v}^*(t)$. \hfill$\blacksquare$
\end{pf}
\subsubsection{The beacons move with the same bounded velocity}
Next, we consider the problem where the beacons are moving at the same unknown bounded time-varying velocity, i.e., $\|\m{v}^*\|_{\infty} \leq \eta$. It is assumed that $\m{v}^*$ is uniformly continuous and the upper bound $\eta$ is available to the agent. We propose the following control law
\begin{equation}\label{eq:bound moving first order}
\begin{aligned}
\m{u}(t) = k\sum_{i=1}^n \omega_i {\m{g}}_i + \beta \text{sgn}\left( \sum_{i=1}^n \omega_i {\m{g}}_i\right),
\end{aligned}
\end{equation}
where the constants $\beta > \eta$ and $k \geq 0$. 
\begin{theorem}\label{thm:bound moving first order}
Let the beacons' velocity $\m{v}^*(t)$ be a bounded uniformly continuous function with a known upper bound $\eta$. If there is no collision among the agent and beacons, under the control law \eqref{eq:bound moving first order}, the tracking error $\bm{\delta}(t)$ is uniformly bounded and $\bm{\delta}(t) \to \m{0}_d$, as $t\to +\infty$.
\end{theorem}
\begin{pf} Since $\m{u}(t)$ is discontinuous, the solution of the system is understood in Filippov sense \cite{Shevitz1994lyapunov}. Consider the function $V = {\m{z}}^\top\bar{\m{W}}({\m{g}}-{\m{g}}^*)$, we have
{\small \begin{align}
\Dot{V} &\in^{\text{a.e}} ({\m{g}}-{\m{g}}^*)^\top\bar{\m{W}}\text{K}[\Dot{{\m{z}}}] \nonumber\\
&= \left(\sum_{i=1}^n\omega_i {\m{g}}_i\right)^\top\left( \m{v}^* - k\sum_{i=1}^n \omega_i {\m{g}}_i - \beta \text{K[sgn]}\left( \sum_{i=1}^n \omega_i {\m{g}}_i\right) \right) \nonumber\\
&= -k\left|\left|\sum_{i=1}^n\omega_i {\m{g}}_i\right|\right|_2^2 - \beta \left|\left|\sum_{i=1}^n\omega_i {\m{g}}_i\right|\right|_1 + \left(\sum_{i=1}^n\omega_i {\m{g}}_i\right)^\top\m{v}^*\nonumber\\
&\leq -k\left|\left|\sum_{i=1}^n\omega_i {\m{g}}_i\right|\right|_2^2 - \beta \left|\left|\sum_{i=1}^n\omega_i {\m{g}}_i\right|\right|_1 + \left|\left|\sum_{i=1}^n\omega_i {\m{g}}_i\right|\right|_1\left|\left|\m{v}^*\right|\right|_\infty\nonumber\\
&\leq -k \left| \left|\sum_{i=1}^n\omega_i {\m{g}}_i\right|\right|_2^2 - (\beta-\eta) \left|\left|\sum_{i=1}^n\omega_i {\m{g}}_i\right|\right|_1 \leq 0.
\end{align}}
It follows that $V(t) \leq V(0)$, $\forall t>0$, and ${\m{z}}^\top\bar{\m{W}}({\m{g}}-{\m{g}}^*)$ is always bounded. 
Suppose that $V(0) = \alpha$, we have $\m{z}^\top\bar{\m{W}}({\m{g}}-{\m{g}}^*) \leq \alpha$, which is equivalent to
\begin{equation*}
\begin{aligned}
\frac{\|\bm{\delta}(t)\|^2}{\big(\|\bm{\delta}(t)\|+\max_{i} {d}_i^*\big)} &\leq \frac{2\alpha}{\lambda_{\min}(\sum_{i = 1}^n \omega_i\mathbf{P}_{{g}^*_i})}= \zeta.
\end{aligned}
\end{equation*}
The above inequality implies $\|\bm{\delta}(t)\| \in [0,\xi]$, where $\xi = (\zeta + \sqrt{\zeta^2+4\zeta\max_{i} {d}_i^*})/2$. Hence, $\bm{\delta}(t)$ is uniformly bounded. As there is no collision, $\Dot{{\m{g}}}_i = \frac{\m{P}_{{\m{g}}_i}}{\hat{d}_i}(\m{v}^*-\m{v})$ is bounded so that $\sum_{i=1}^n\omega_i {\m{g}}_i(t)$ is bounded and uniformly continuous. As a result, $\dot{V}$ is also uniformly continuous. 
From Barbalat's lemma \cite{Khalil2002nonlinear}, $\dot{V}\rightarrow 0$ as $t \rightarrow \infty$, i.e., $\bm{\delta} = \m{p} -{\m{p}}^* \to \m{0}_d$, as $t \rightarrow \infty$.\hfill $\blacksquare$
\end{pf}
In practice, the upper bound of beacons' velocity $\eta$ may not be known by the agent. 
To this end, we may relax the objective from perfectly  tracking to maintaining in a close neighborhood of the Fermat-Weber point. Inspired of adaptive sliding mode control design \cite{Roy2020adaptive}, the following control law is proposed 
\begin{equation}\label{eq:adaptive moving approx first order}
\begin{aligned}
\m{u}(t) &= k\sum_{i=1}^n \omega_i {\m{g}}_i + \beta \text{sgn}\left( \sum_{i=1}^n \omega_i{\m{g}}_i\right),\\
\Dot{\beta}(t) &= k_{\beta}\left(\left|\left|\sum_{i=1}^n\omega_i {\m{g}}_i\right|\right|_1 - \tau_{\beta}\beta\right), 
\end{aligned}
\end{equation}
where $k_{\beta},\tau_\beta$ are positive constants.

 \begin{theorem}\label{thm:SMC_linkage}
Suppose there is no collision between the agent and beacons, and $\mathbf{p}(0) \in  \mc{B}_{R}$. Under the control law (\ref{eq:adaptive moving approx first order}), the agent will converge to a small neighborhood of the actual Fermat-Weber point after a finite time.
\end{theorem}
\begin{pf} 
Consider the following Lyapunov function
$V = {\m{z}}^\top\bar{\m{W}}({\m{g}}-{\m{g}}^*) + \frac{1}{2k_\beta}(\beta - \overline{\beta})^2,$ where $\overline{\beta} > \eta$. The derivative of $V$ can be obtained as follows
\begin{align}
\Dot{V} &= \Big(\sum_{i=1}^n\omega_i {\m{g}}_i\Big)^\top\left( \m{v}^* - k\sum_{i=1}^n \omega_i {\m{g}}_i - \beta \text{sgn}\Big(\sum_{i=1}^n \omega_i {\m{g}}_i\Big) \right) \nonumber\\
&\qquad + (\beta - \overline{\beta})\left(\left|\left|\sum_{i=1}^n\omega_i {\m{g}}_i\right|\right|_1 - \tau_{\beta}\beta\right) \nonumber\\
\leq& -k\left|\left|\sum_{i=1}^n\omega_i {\m{g}}_i\right|\right|^2 -\overline{\beta} \left|\left|\sum_{i=1}^n\omega_i {\m{g}}_i\right|\right|_1 + \left|\left|\sum_{i=1}^n\omega_i {\m{g}}_i\right|\right|_1\|\m{v}^*\|_\infty \nonumber\\
&\qquad -\frac{\tau_\beta}{2}(2\beta^2-2\beta \overline{\beta}+\overline{\beta}^2)+\frac{\tau_\beta}{2}\overline{\beta}^2 \nonumber\\
\leq& -k\left|\left|\sum_{i=1}^n\omega_i ({\m{g}}_i-{\m{g}}_i^*)\right|\right|^2 - \frac{\tau_\beta}{2}(\beta-\overline{\beta})^2+\frac{\tau_\beta}{2}\overline{\beta}^2.\label{eq:SMC_linkage}
\end{align}
Similar to the proof of Theorem \ref{thm:finite time}, we have
\begin{equation}
\begin{aligned}
    \dot{V} &\leq -\underbrace{\min \big(k\chi h(R) ,k_{\beta}\tau_{\beta}\big)}_{\varrho > 0}V +\frac{\tau_\beta}{2}\overline{\beta}^2\\
    &=-\varrho (1-\mu) V - \varrho \mu V + \frac{\tau_\beta}{2}\overline{\beta}^2,
\end{aligned}
\end{equation}
where $\mu \in (0,1).$ It is clearly seen that the time derivative of $V$ is negative outside of the compact set $\mc{D} = \{\|\bm{\delta}(t)\| < R|~ V(t) \leq \frac{\tau_\beta\overline{\beta}^2}{2\varrho \mu}\}$. In the set $\mc{D}$, Corollary \ref{cor:1a} implies the following
\begin{equation*}
\begin{aligned}
\frac{\|\bm{\delta}(t)\|^2}{\|\bm{\delta}(t)\|+\max_{i} {d}_i^*} \leq \frac{\tau_\beta\overline{\beta}^2}{\lambda_{\min}(\sum_{i = 1}^n \omega_i\mathbf{P}_{{g}^*_i})\varrho \mu}= \zeta,
\end{aligned}
\end{equation*}
which is equivalent to $\|\bm{\delta}(t)\| \in [0,\xi]$, where $\xi = (\zeta + \sqrt{\zeta^2+4\zeta\max_{i} {d}_i^*})/2$. Thus, for all $\m{p} \in \mc{B}_R  \setminus \mc{D}$, the equality $\dot{V} \leq -\varrho (1-\mu) V$ holds. This implies that $V$ will decrease until $\m{p}(t) \in \mc{B}_{\xi}$ within a finite time. Finally, the ultimate bound $\xi$ can be selected sufficiently small given that $k, k_{\beta}\tau_{\beta}$ are selected large enough. This completes the proof of this theorem. \hfill $\blacksquare$
\end{pf}

\begin{lemma}\label{collision avoidance}
    Under the control laws (\ref{eq:finite_time}),(\ref{eq:PI-control}),(\ref{eq:bound moving first order}) or (\ref{eq:adaptive moving approx first order}), there always exists a small positive constant $\alpha$ such that if the Lyapunov functions satisfy $V(0) \leq \alpha$, the collision can be avoided.
\end{lemma}
\begin{pf} Similar to the proof of Theorem \ref{thm:bound moving first order}, the inequality $V(t) \leq \alpha$ implies  $\|\bm{\delta}(t)\| \in [0,\xi]$, where $\xi = (\zeta + \sqrt{\zeta^2+4\zeta\max_{i} {d}_i^*})/2$. Hence, once can always choose a small $\alpha$ so that $\xi < \min_i d_i^*$. Hence, $\bm{\delta}(t) < \min_i d_i^*$, i.e., $\bm{p}(t) \in \mc{B}_R$. At time $t = 0$, similar to Remark \ref{remark:collision_first_stationary}, since $\bm{\delta}(0) < \min_i d_i^*$, $d_i > \epsilon$ and there is no collision initially. As $\dot{V} \leq 0$, $\|\bm{\delta(t)}\|$ cannot escape $[0, \xi]$, and the collision avoidance is guaranteed. Otherwise, there must be an instant $t_1$ such that $V(t_1) = \alpha$ and $\dot{V}(t_1) > 0$, which is impossible. \hfill $\blacksquare$
\end{pf}
\section{Double-Integrator Agent}\label{sec: Double-Integrator} 
\subsection{Stationary Beacons}\label{subsec: Double-Integrator Stationary} 
Consider the case where agent's dynamic is modeled as (\ref{eq:double}), and beacons are stationary. The following PD-like bearing-only control law is proposed
\begin{align} \label{eq:PD}
\m{u} &= \sum_{i=1}^n \omega_i {\m{g}}_i - k\m{v}, \, \text{with } k>0.
\end{align}
\begin{theorem}\label{thm:1a}
Suppose there is no collision and the agent can obtain its own velocity, under the control law \eqref{eq:PD}, for all $\m{p}(0)=[p_{1}(0),\ldots,p_{d}(0)]^\top$, $\m{p}(t)$ asymptotically converges to ${\m{p}}^*$. 
\end{theorem}
\begin{pf} Consider the Lyapunov function \[V = {\m{z}}^\top\bar{\m{W}}({\m{g}}-{\m{g}}^*) + \frac{\mathbf{v}^\top \mathbf{v}}{2}.\] The derivative of $V$ is as follows
\begin{equation}
\begin{aligned}
\Dot{V} &= ({\m{g}}-{\m{g}}^*)^\top\bar{\m{W}}\Dot{{\m{z}}} + \m{v}^\top \dot{\m{v}}\\
&= -\left(\sum_{i=1}^n\omega_i {\m{g}}_i\right)^\top\m{v} +\m{v}^\top \left( \sum_{i=1}^n \omega_i {\m{g}}_i - k\m{v}\right)\\
&= -k\|\m{v}\|^2 \leq 0. 
\end{aligned}
\end{equation} Since $V \ge 0$ and $\dot{V}\leq 0$, $V(t)$ is always bounded, so does $\|\m{v}(t)\|$. Based on the LaSalle's invariance principle, each trajectory of the system asymptotically converges to the largest invariant set inside $\Omega = \{\dot{V}=0\}$, i.e., $\m{v}\equiv \m{0}_{d}$. Whenever $\m{v} \equiv \m{0}_{d}$, it is required that $\dot{\m{v}} \equiv \m{0}_{d}$. From equation  \eqref{eq:PD}, it follows that $\sum_{i=1}^n \omega_i {\m{g}}_i \equiv \m{0}_{d}$. This implies that ${\m{g}}_i \equiv \m{g}_i^*$ and $\m{p}(t) \equiv {\m{p}}^*$. Thus, $\m{p}(t)\to {\m{p}}^*$ and $\m{v}(t) \to \m{0}_{d}$ as $t\to \infty$.
\hfill$\blacksquare$
\end{pf}
\subsection{Moving Beacons}\label{sec: Double-Integrator Moving} 
\subsubsection{The beacons move with the same constant velocity}\label{subsec: Double-Integrator Constant} 
Let the beacons move with unknown constant velocity, i.e., $\dot{\m{v}}^* = \m{0}_{d}$. The following adaptive control law is proposed
\begin{equation}
\begin{aligned}\label{eq:PD adaptive}
\m{u} &= (k_2+1)\sum_{i=1}^n \omega_i {\m{g}}_i - k_1(\m{v}-\hat{\m{v}}),\\
\dot{\hat{\m{v}}} &= k_2\sum_{i=1}^n \omega_i {\m{g}}_i,\\
\end{aligned}
\end{equation}
where $k_1,k_2>0$. The following theorem yields
\begin{theorem}\label{thm:constant_2_order}
Suppose that there is no collision. Under the control law \eqref{eq:PD adaptive}, for all initial conditions, $\m{p}(t) \to {\m{p}}^*(t)$ and ${\m{v}}(t) \rightarrow \m{v}^*$ as $t \rightarrow \infty$.
\end{theorem}
\begin{pf} Consider the following Lyapunov function: \[V = {\m{z}}^\top\bar{\m{W}}({\m{g}}-{\m{g}}^*) + \frac{1}{2}\|\m{v}-\hat{\m{v}}\|^2 + \frac{1}{2k_2}\|\hat{\m{v}}-\m{v}^*\|^2.\]
The derivative of $V$ can be obtained as follows
\begin{equation}
\begin{aligned}
\dot{V} =& - \left(\sum_{i=1}^n \omega_i {\m{g}}_i\right)^\top (\m{v}-\hat{\m{v}}+\hat{\m{v}}-\m{v}^*)\\
&\quad +(\m{v}-\hat{\m{v}})^\top(\m{u-\Dot{\hat{\m{v}}}})
+(\hat{\m{v}}-\m{v}^*)^\top\frac{\Dot{\hat{\m{v}}}}{k_2}\\
=&-k_1\|\m{v}-\hat{\m{v}}\|^2\leq 0 .
\end{aligned}
\end{equation}
Based on the LaSalle's invariance principle, each trajectory of the system asymptotically converges to the largest invariant set inside $\Omega = \{\dot{V}=0\}$, i.e., $\m{v}(t)=\hat{\m{v}}(t)$. For $\m{v}\equiv \hat{\m{v}}$, we also have $\m{u}\equiv \dot{\hat{\m{v}}}$. From  \eqref{eq:PD adaptive}, it follows that $\sum_{i=1}^n \omega_i {\m{g}}_i \equiv \m{0}_{d+1}$. This implies that ${\m{g}}_i \equiv \m{g}_i^*$ and $\m{p}(t) \equiv {\m{p}}^*$. Thus, $\m{p}(t)\to {\m{p}}^*$ and $\m{v}(t) \to \m{v}^*$ as $t\to \infty$.
\hfill$\blacksquare$
\end{pf}

\subsubsection{The beacons move with the same uniformly bounded velocity}\label{subsec: Double-Integrator Bounded} 

We next consider the case where the beacons move with unknown uniformly bounded velocity $\m{v}^*(t)$. Since $\m{v}^*$ is assumed to be uniformly bounded, its derivative $\dot{\m{v}}^*(t)$ is also bounded, i.e., $\|\dot{\m{v}}^*(t)\|_{\infty} \leq \eta $. Suppose that the relative velocity $\m{v}(t)-\m{v}^*(t)$ is measurable, the following bearing-based control law is introduced
\begin{equation}\label{eq:moving 2nd order}
\begin{aligned}
\dot{\m{q}} =& \sum_{i=1}^n\omega_i{\m{g}}_i-\m{v}+\m{v}^*,~\m{r} = \m{q} - \m{v} + \m{v}^*,\\
\m{u} =& 2\sum_{i=1}^n\omega_i{\m{g}}_i-2(\m{v}-\m{v}^*) + \beta \text{sgn}(\m{r}),
\end{aligned}
\end{equation}
where $\beta > \eta$. 
\begin{theorem}\label{thm: moving 2nd order}
Suppose that the relative velocity and the upper bound of the beacons' velocity are available. If there is no collision, under the control law (\ref{eq:moving 2nd order}), $\m{p}(t) \to {\m{p}}^*(t)$, as $t \to +\infty$.
\end{theorem}
\begin{pf} Consider the following Lyapunov function \[V =  {\m{z}}^\top\bar{\m{W}}({\m{g}}-{\m{g}}^*) + \frac{\|\m{q}\|^2}{2}+ \frac{\|\m{q}-\m{v}+\m{v}^*\|^2}{2},\] we have
\begin{align}
    \dot{V} =&  -(\m{v}-\m{v}^*)^\top\sum_{i=1}^n\omega_i{\m{g}}_i + \m{q}^\top(\sum_{i=1}^n\omega_i{\m{g}}_i-\m{v}+\m{v}^*) \nonumber\\
    &\qquad + (\m{q}-\m{v}+\m{v}^*)^\top(\Dot{\m{q}}-\m{u}+\Dot{\m{v}}^*) \nonumber\\ 
    =&-\|\m{v}-\m{v}^*\|^2 - \beta \|\m{q}-\m{v}+\m{v}^*\|_1 \nonumber\\
    &\qquad +(\m{q} -\m{v}+\m{v}^*)^\top\dot{\m{v}}^*(t) \nonumber\\
    \leq& -\|\m{v}-\m{v}^*\|^2 -(\beta - \eta)\|\m{q}-\m{v}+\m{v}^*\|_1 \nonumber\\
    \leq & 0.
\end{align}
Hence, it follows that $\m{v} \rightarrow \m{v}^*$ and $\m{q} \rightarrow \m{0}$ as $t \rightarrow \infty$. Since $V > 0$ and $\Dot{V} \leq 0$, $V$ is bounded, so do $\m{q},\m{v}$ and $\m{u}$. As there is no collision, $\Dot{{\m{g}}}_i = \frac{\m{P}_{{\m{g}}_i}}{\hat{d}_i}(\m{v}^*-\m{v})$ is bounded. Thus, it can be seen that $\Ddot{\m{q}} = \sum_{i=1}^n\omega_i\dot{{\m{g}}}_i -\m{u}+\Dot{\m{v}}^*$ is also bounded. According to the Barbalat’s lemma, $\Dot{\m{q}}\rightarrow \m{0}$ as $t \rightarrow \infty$. Combining with the first equation of (\ref{eq:moving 2nd order}), $\sum_{i=1}^n\omega_i{\m{g}}_i \rightarrow 0$ as $t \rightarrow \infty$. This completes the proof of Theorem \ref{thm: moving 2nd order}. \hfill$\blacksquare$
\end{pf}

\begin{figure}[t]
    \centering
\subfloat[]{\includegraphics[width=0.48\linewidth]{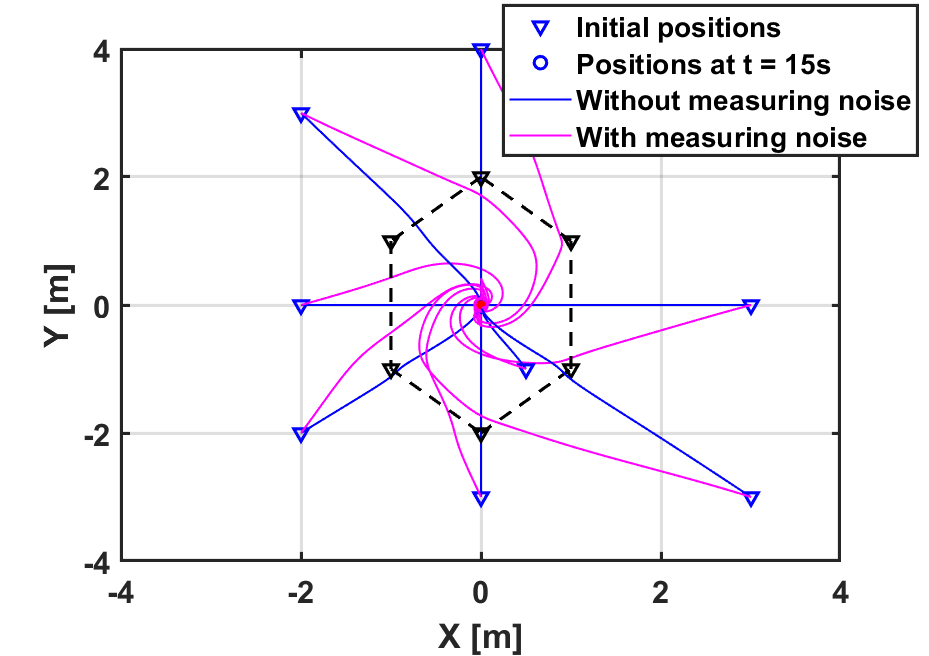}}
\hfill
\subfloat[]{\includegraphics[width=0.48\linewidth]{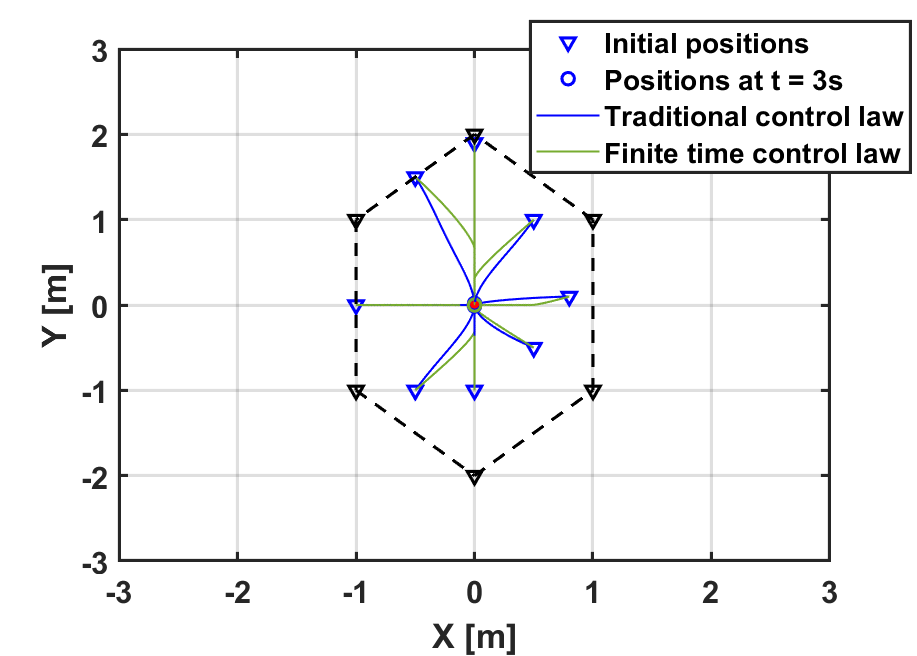}}
\\
\subfloat[]{\includegraphics[width=0.48\linewidth]{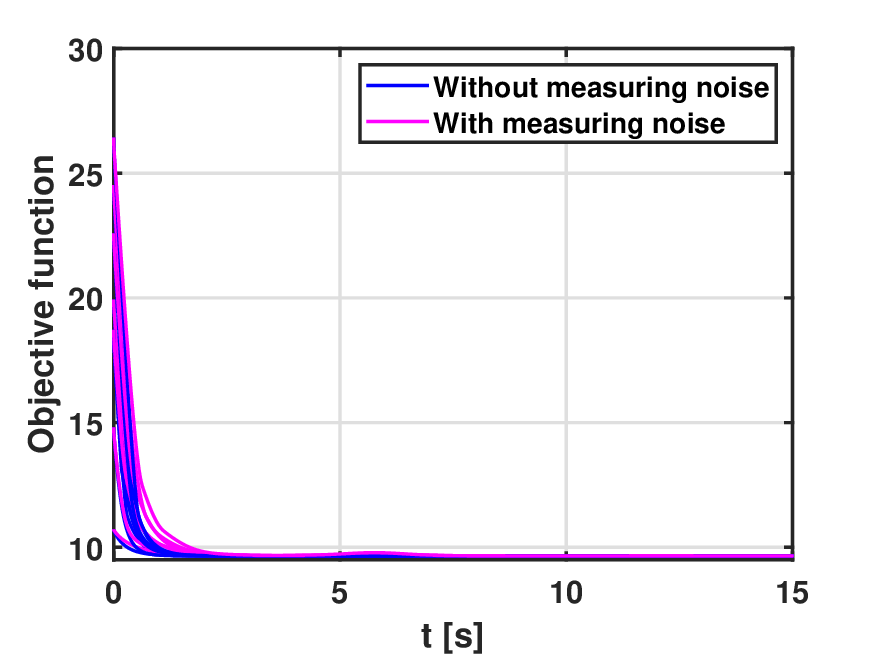}}
\hfill
\subfloat[]{\includegraphics[width=0.48\linewidth]{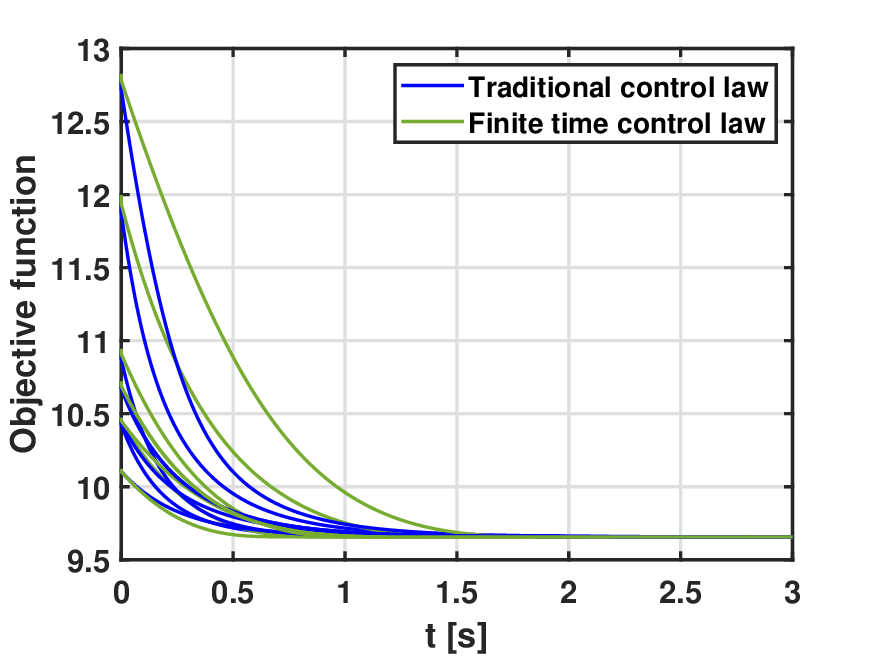}}
\hfill\\
\subfloat[]{\includegraphics[width=0.48\linewidth]{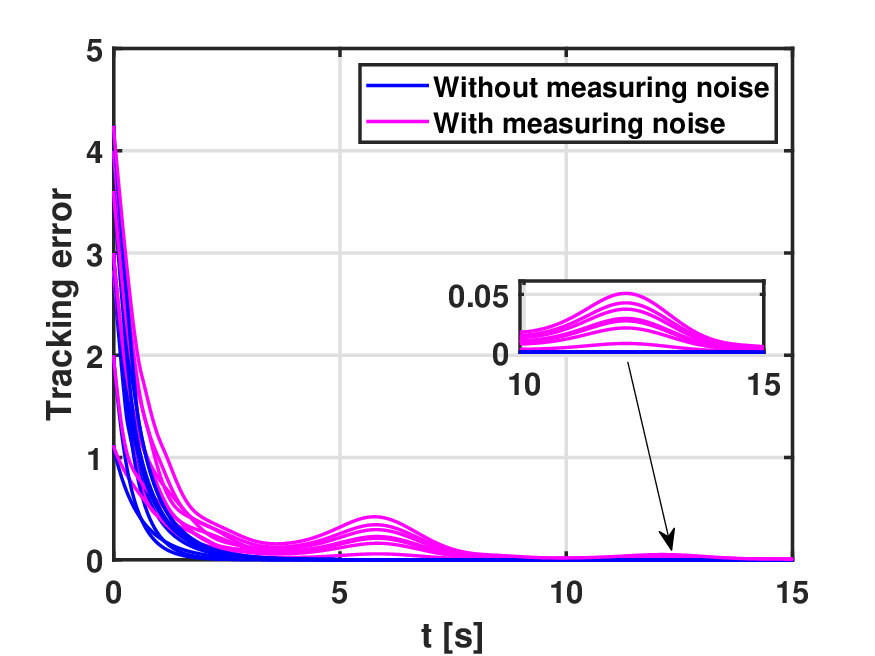}}
\hfill
\subfloat[]{\includegraphics[width=0.48\linewidth]{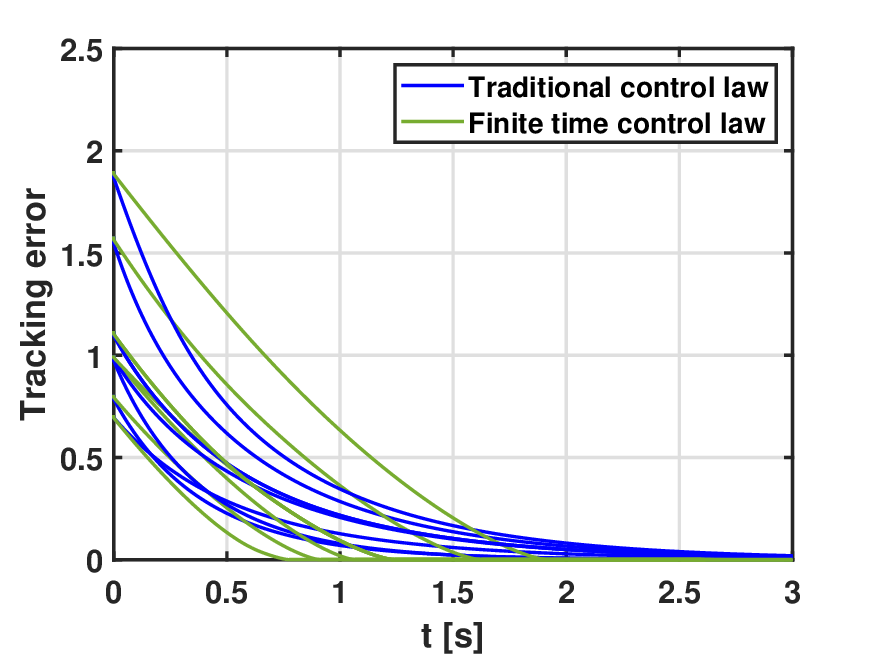}}
\caption{Simulation 1a: Single-integrator agent model and stationary beacons. Gradient control law (\ref{eq:control_law_gradient_descent}) with and without measuring noise (a)-(c)-(e), and comparison with finite time control law (\ref{eq:finite_time}) (b)-(d)-(f). (a) and (b) Trajectories of the agent, beacons, and the Fermat-Weber point. The initial and final positions of agents are marked with $'\triangledown'$ and $'\circ'$, respectively; (c) and (d) The  function $f(\m{p})$ versus time; (e) and (f) The norm of tracking error $\|\bm{\delta}(t)\|$ versus time.} \label{fig:single_stationary}
\end{figure}


\section{Simulation Results}\label{sec:simulation}
This section presents a set of simulation examples to illustrate the theoretical results. 

\subsection{Simulation 1: Single-integrator agent model}
We conduct this simulation in two-dimensional space. There are six beacons, whose positions at the time $t = 0$ are:
$\m{p}_1(0) = [1, 1]^\top, \m{p}_2(0) = [0,2]^\top, \m{p}_3(0) = [-1, 1]^\top, \m{p}_4(0) = [-1,-1]^\top, \m{p}_5(0) = [0,-2]^\top,  \m{p}_6(0) = [1,-1]^\top$. The positive constants $\omega_i$ are chosen as $\omega_1 = \dots=\omega_6 = 1$. Thus, the Fermat-Weber point $\m{p}^*$ is actually the average value of beacons' positions. 
 
\begin{figure}[t]
\centering
\subfloat[]{\includegraphics[width=0.48\linewidth]{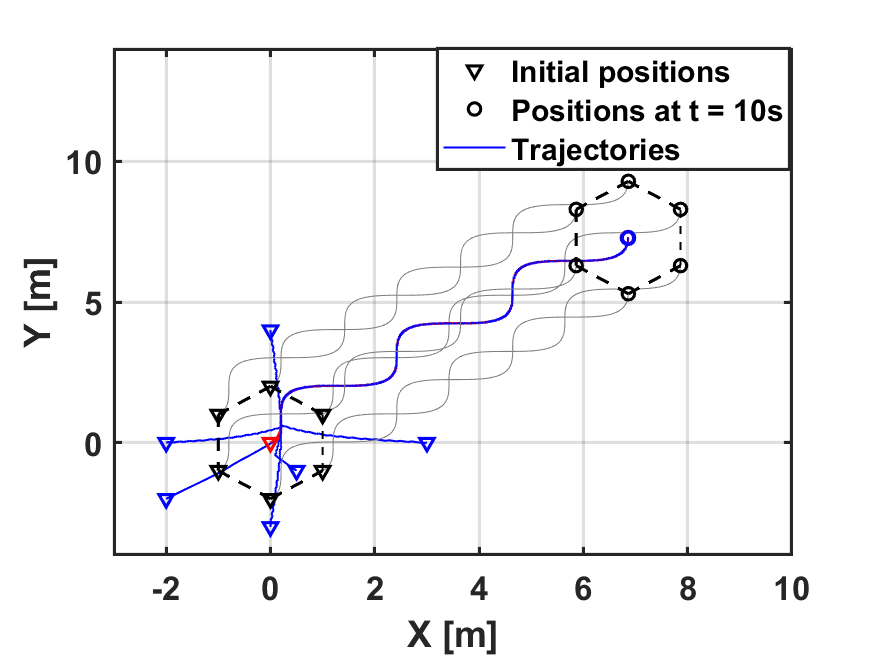}}
\hfill
\subfloat[]{\includegraphics[width=0.48\linewidth]{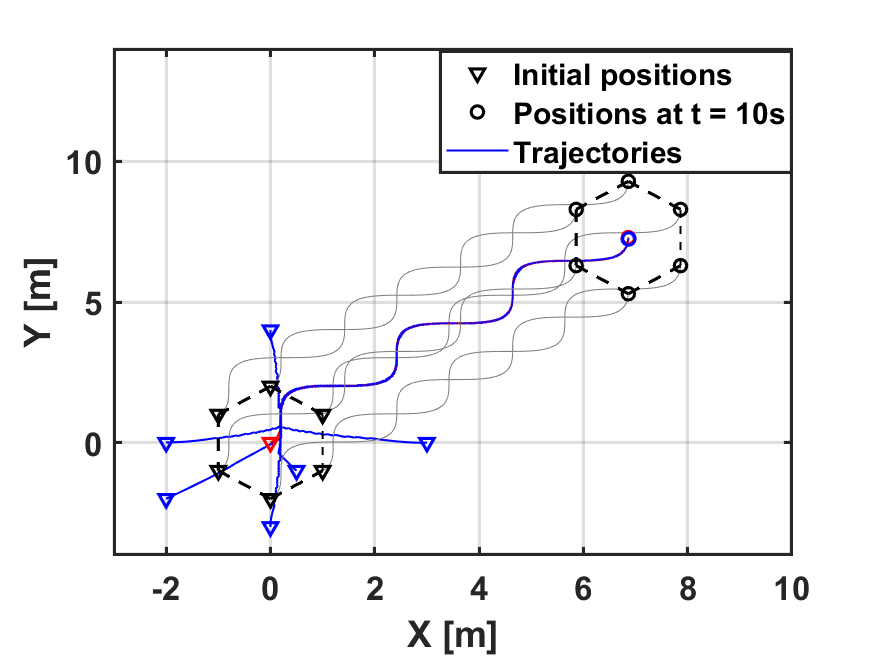}}
\\
\subfloat[]{\includegraphics[width=0.48\linewidth]{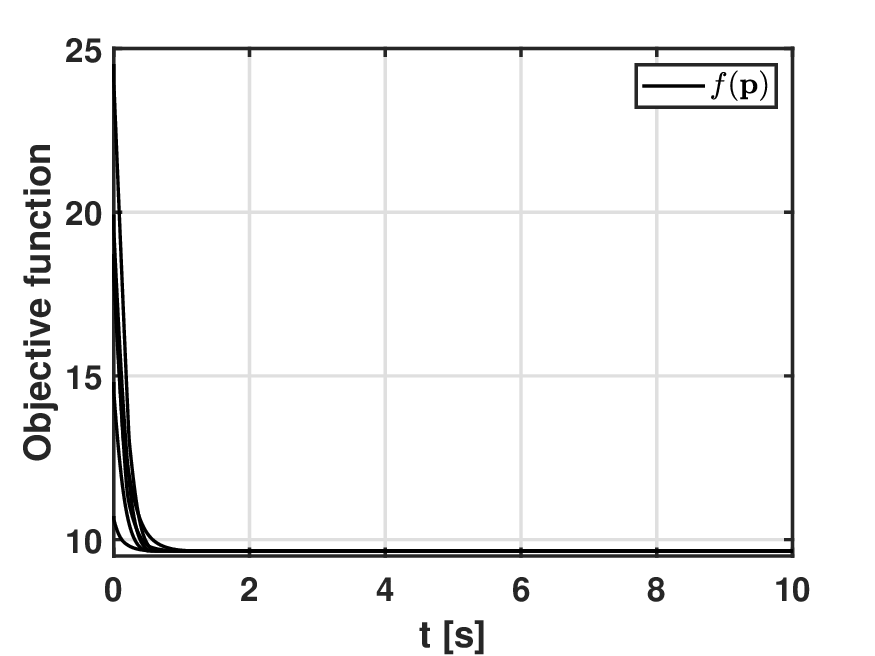}}
\hfill
\subfloat[]{\includegraphics[width=0.48\linewidth]{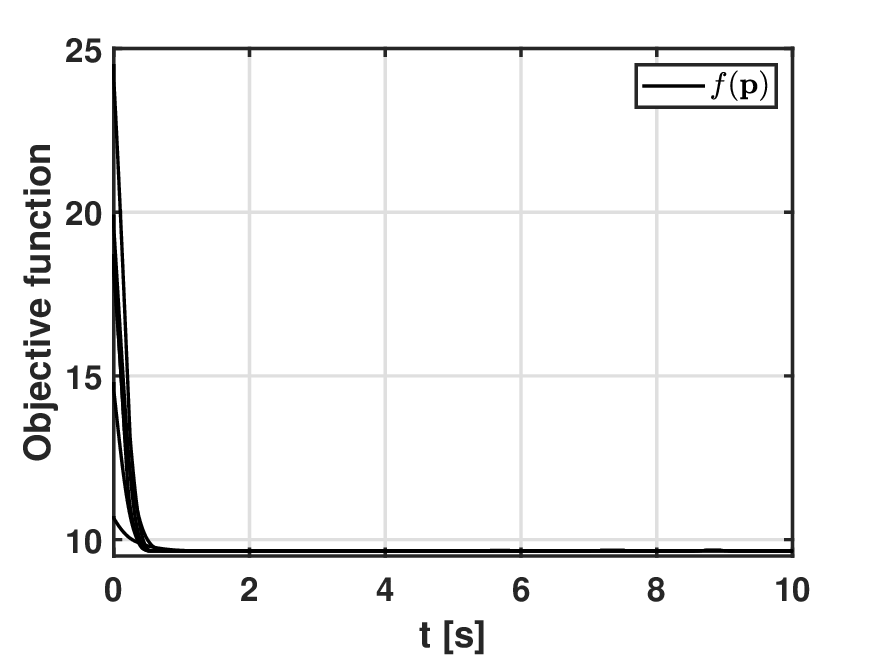}}\\
\subfloat[]{\includegraphics[width=0.48\linewidth]{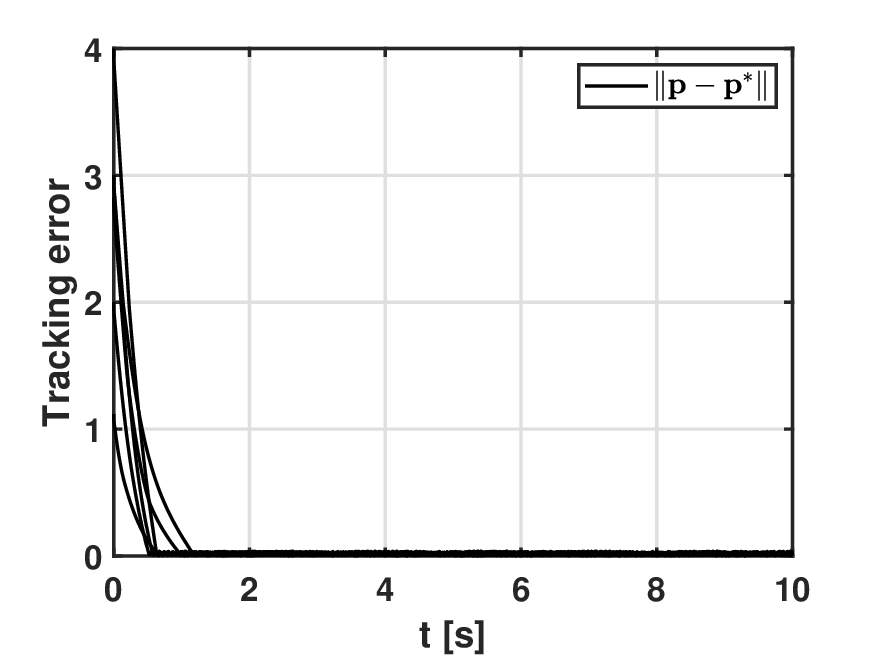}}
\hfill
\subfloat[]{\includegraphics[width=0.48\linewidth]{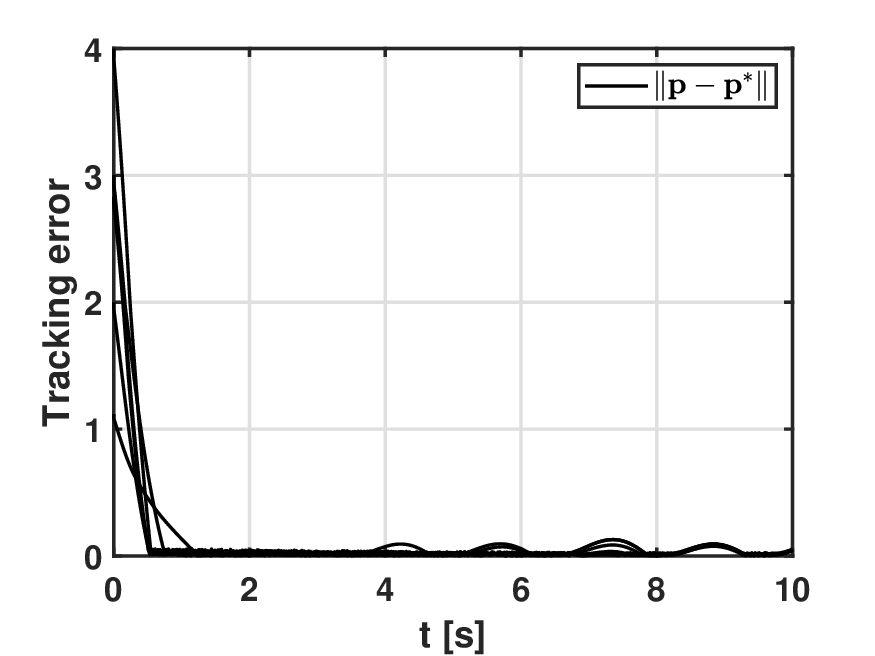}}
\caption{Simulation 1c: Single-integrator agent model and moving beacons with time-varying velocity under the control laws (\ref{eq:bound moving first order}) (a)-(c)-(e) and (\ref{eq:adaptive moving approx first order}) (b)-(d)-(f).   (a) and (b) The trajectories of the agent, beacons, and Fermat-Weber point. Their positions at $t = 0$ and $t = 10$ sec are marked with $'\triangledown'$ and $'\circ'$, respectively; (c) and (d) The  function $f(\m{p})$ versus time; (e) and (f) The magnitude of the tracking error $\|\bm{\delta}(t)\|$ versus time. }\label{fig:single_time}
\end{figure}

\textit{Simulation 1a - Stationary beacons:} Consider the scenario where beacons are stationary. Fig. \ref{fig:single_time}a--\ref{fig:single_time}c demonstrate the trajectories of the agent with different initial positions under gradient control law (\ref{eq:control_law_gradient_descent}), with and without disturbance. The measuring errors are chosen as $\theta_1(t)=\theta_4(t) = \frac{\pi}{3}$, $\theta_2(t)=\theta_5(t)=\sin(t)$ and $\theta_3(t)=\theta_6(t)=5\sin(t)/4$. In addition, Fig. \ref{fig:single_stationary}d--\ref{fig:single_stationary}f show the comparison between the traditional control law and the proposed finite-time control law in (\ref{eq:finite_time}). The coefficient of the controller is chosen as $a = 0.3$. To align with the theoretical results, the agent's initial position is randomly chosen to lie within $\mc{B}_R$. 
\begin{figure}[t]
\centering
\subfloat[]{\includegraphics[width=0.8\linewidth]{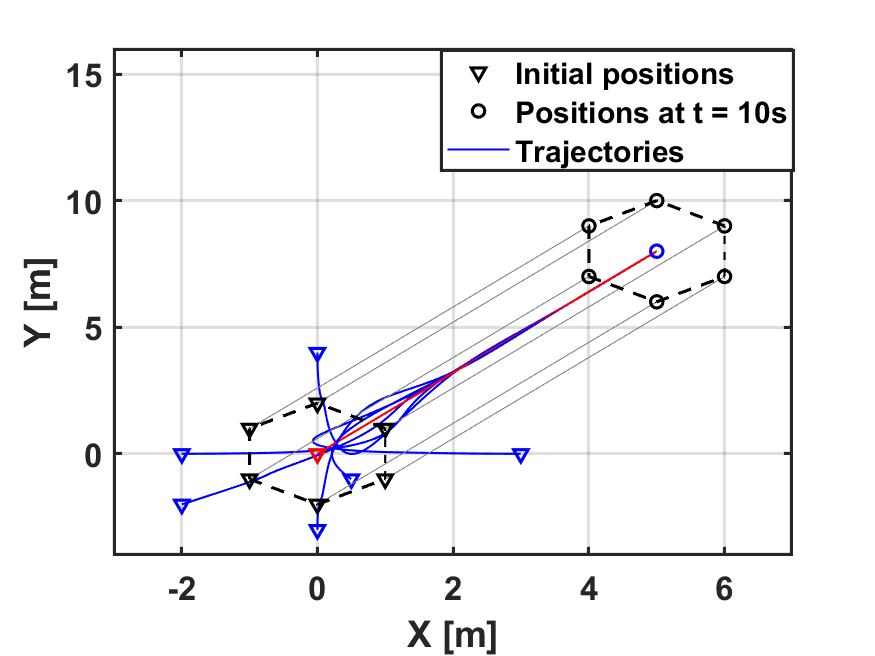}}\label{fig:single_const_trajectories}
\\
\subfloat[]{\includegraphics[width=0.48\linewidth]{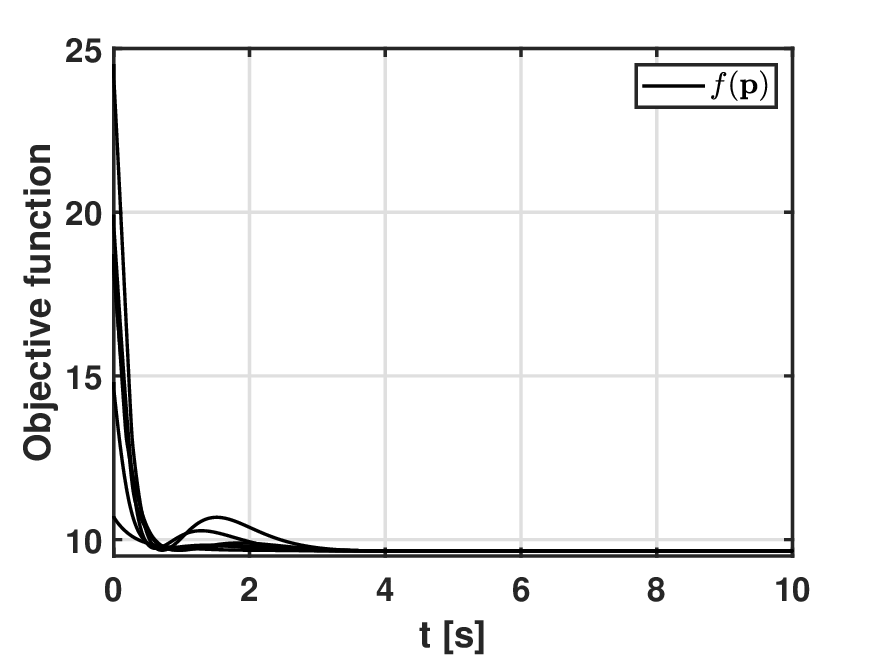}}\label{fig:single_const_cost}
\hfill
\subfloat[]{\includegraphics[width=0.48\linewidth]{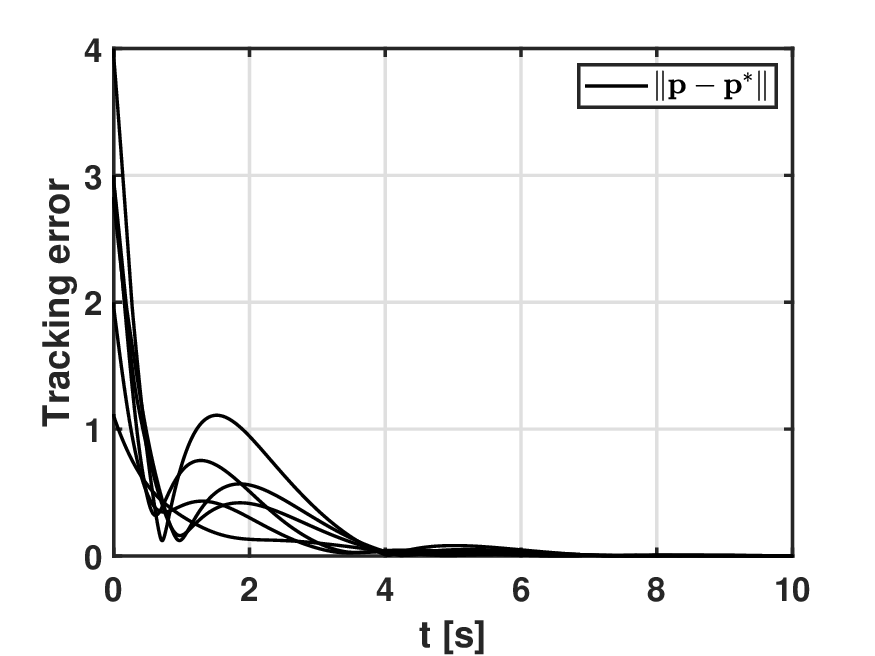}}\label{fig:single_const_error}
\caption{Simulation 1b: Single-integrator agent model and moving beacons with constant velocity:  (a) The trajectories of the agent, beacons, and Fermat-Weber point are colored blue, gray, and red, respectively. Their positions at $t = 0$ and $t = 10$ sec are marked with $'\triangledown'$ and $'\circ'$, respectively; (b) The  function $f(\m{p})$ versus time; (c) The magnitude of the tracking error $\|\bm{\delta}(t)\|$ versus time. }\label{fig:single_const}
\end{figure}

As can be seen from Fig. \ref{fig:single_stationary}a--\ref{fig:single_stationary}c, the measuring error deviates the agent from reaching $\m{p}^*$. As $\theta_i(t) < \pi/2$, the agent's trajectory stays inside a neighborhood of the  Fermat-Weber point. Meanwhile, it is clearly seen that the finite-time control law provides higher convergence rate than the traditional gradient descent control law if the agent is initialized inside $\mc{B}_R$.

\textit{Simulation 1b - Beacons move with constant velocity:} Suppose that beacons move with constant velocity $\m{v}^* = 
\begin{bmatrix}
    0.5 &
    0.8
\end{bmatrix}^\top
$ (m/s).
The parameter of the control law (\ref{eq:PI-control}) is chosen as $k=1$. The simulation result after $10s$ is illustrated in Fig. \ref{fig:single_const}a--\ref{fig:single_const}c. It can be seen from Fig.\ref{fig:single_const} (c) that, with different initial positions $\m{p}(0)$, the agent can eventually track the Fermat-Weber point $\m{p}^*(t)$.
\begin{figure*}[t]
\centering
\subfloat[]{\includegraphics[width=0.45\linewidth]{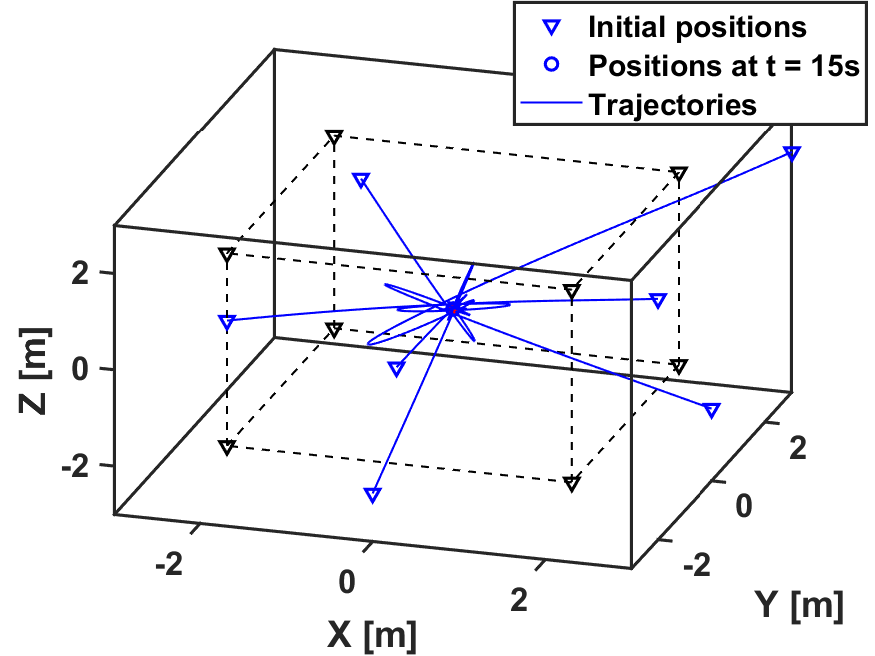}}
\\
\subfloat[]{\includegraphics[width=0.32\linewidth]{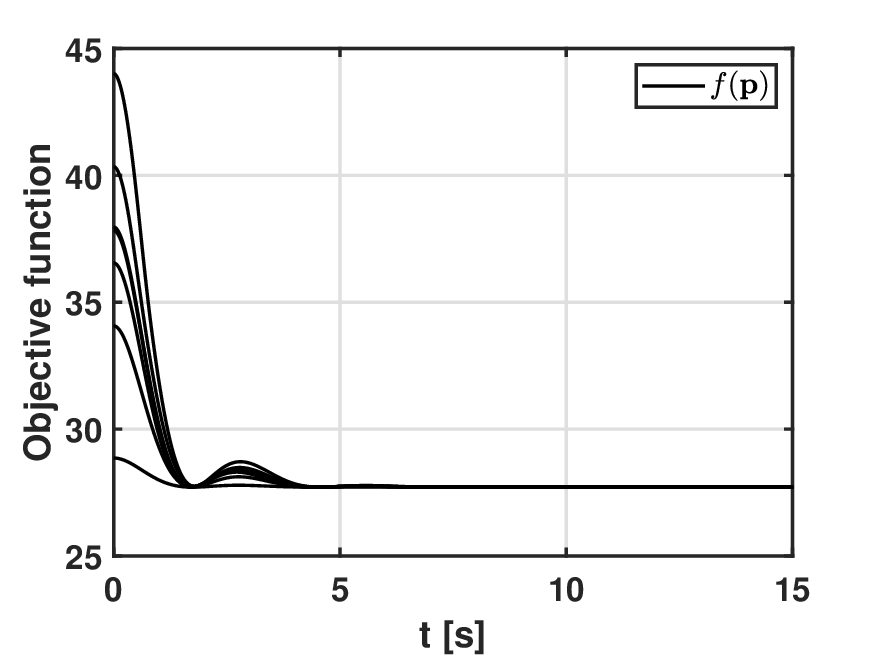}}
\hfill
\subfloat[]{\includegraphics[width=0.32\linewidth]{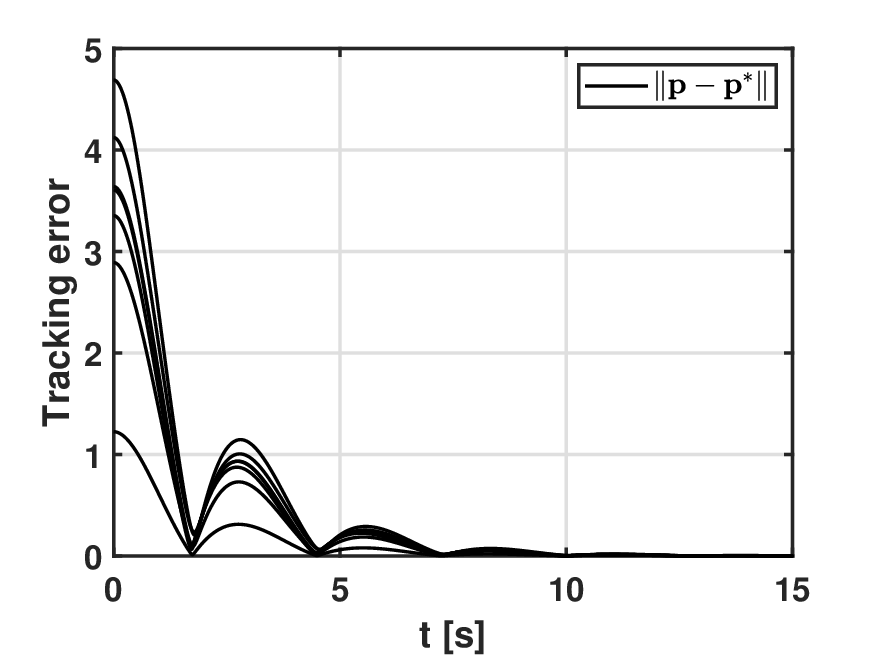}}
\hfill
\subfloat[]{\includegraphics[width=0.32\linewidth]{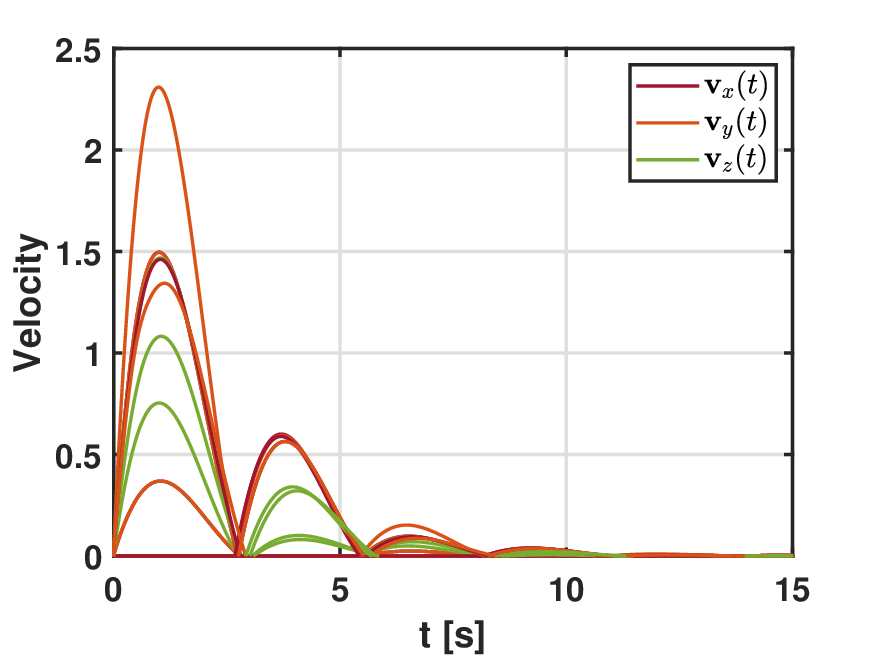}}
\caption{Simulation 2a: Double-integrator agent model and stationary beacons.  (a) The trajectories of the agents. The positions of the beacons, the agent, and the Fermat-Weber point are colored in black, blue, and red, respectively. Their positions at $t = 0$ and $t = 15$ sec are marked with $'\triangledown'$ and $'\circ'$, respectively; (b) The  function $f(\m{p})$ versus time; (c) The norm of tracking error $\|\bm{\delta}(t)\|$ versus time; (d) The velocities of the agent along the x, y, and z axes. }\label{fig:double_stationary}
\end{figure*}

\begin{figure*}[t]
    \centering
\subfloat[]{\includegraphics[width=0.4\linewidth]{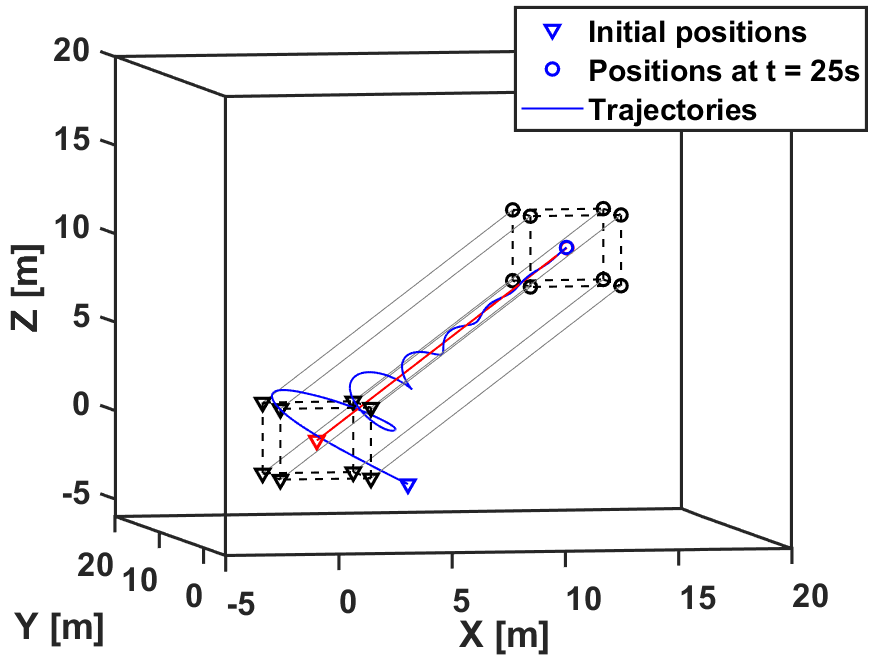}}
\\
\subfloat[]{\includegraphics[width=0.32\linewidth]{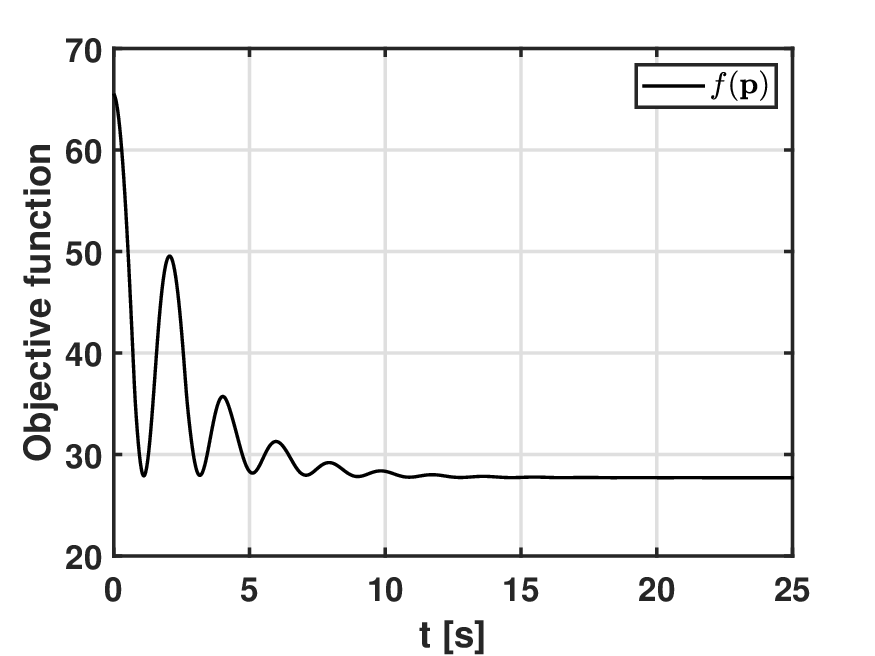}}
\hfill
\subfloat[]{\includegraphics[width=0.32\linewidth]{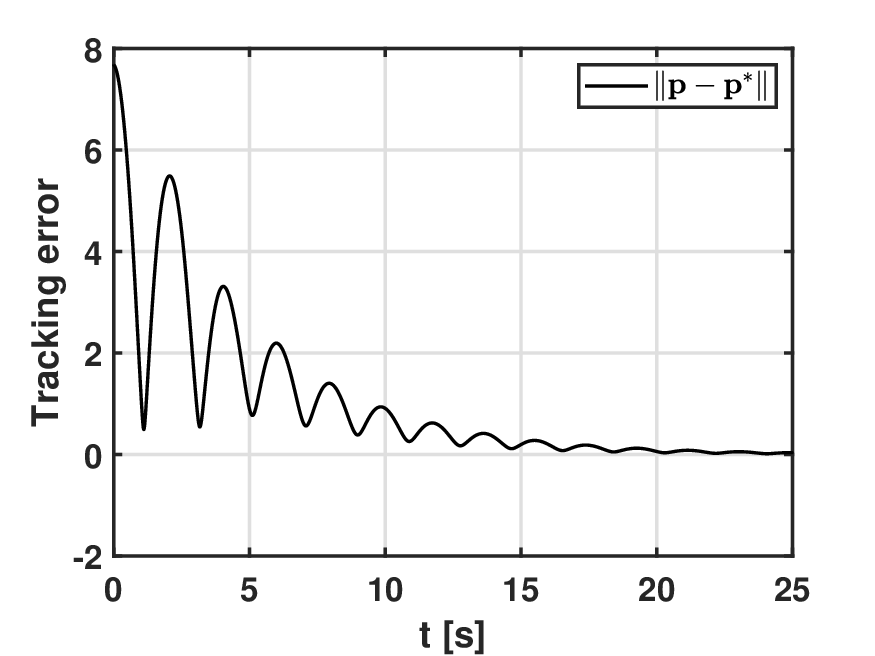}}
\hfill
\subfloat[]{\includegraphics[width=0.32\linewidth]{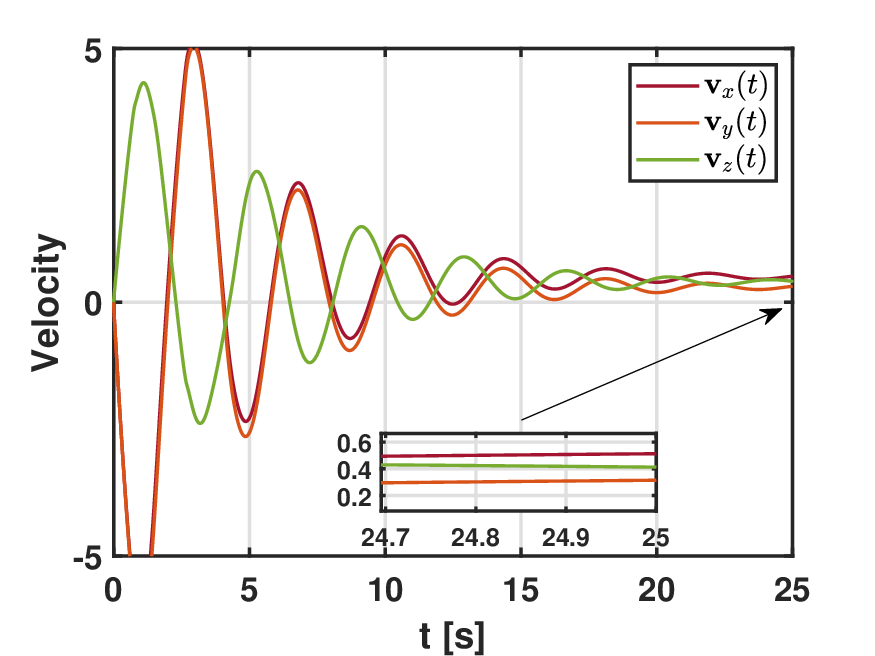}}
\caption{Simulation 2b: Double-integrator agent model and moving beacons with constant velocity.  (a) The trajectories of the beacons, the agent, and the Fermat-Weber point are colored in black, blue, and red, respectively. Their positions at $t = 0$ and $t = 25$ sec are marked with $'\triangledown'$ and $'\circ'$, respectively; (b) The  function $f(\m{p})$ versus time; (c) The norm of tracking error $\|\bm{\delta}(t)\|$ versus time; (d) The velocities of the agent along the x, y, and z axes. }\label{fig:double_constant}
\end{figure*}

\textit{Simulation 1c - Beacons move with time-varying velocity:} Next, we consider the case that the beacons' velocity is time-varying and bounded as follows 
$\m{v}^*(t) = 
\begin{bmatrix}
    &\frac{1}{\sqrt{2}}(1 -\text{sin}(2t))\\
&\frac{1}{\sqrt{2}}(1 + \text{sin}(2t))
\end{bmatrix}$ (m/s). Two control laws (\ref{eq:bound moving first order}) and (\ref{eq:adaptive moving approx first order}) are considered. In (\ref{eq:bound moving first order}), the parameters are chosen as $ k=1, \beta = 2$. Regarding (\ref{eq:adaptive moving approx first order}), we choose $ k=1, k_\beta =2, \tau_\beta = 0.1,$ and $\beta(0) = 1$.
The simulation result after $10s$ is illustrated in Fig. \ref{fig:single_time}a--\ref{fig:single_time}f. Overall, with different initial positions $\m{p}(0)$, controller (\ref{eq:bound moving first order}) can steer the agent to the Fermat-Weber point $\m{p}^*(t)$ asymptotically. Meanwhile, under control law specified in (\ref{eq:adaptive moving approx first order}), the agent ultimately converges to a closed region of $\m{p}^*(t)$.

\subsection{Simulation 2: Double-integrator agent model}
This simulation is conducted in three-dimensional space. At $t = 0$, the positions of the beacons are selected so that they form the vertices of a cube. 
 The positive constants $\omega_i$ are chosen as $\omega_1 = \dots=\omega_8 = 1$. Hence, the Fermat-Weber point is the center of the cube.

\textit{Simulation 2a - Stationary beacons:} Consider the scenario where the beacons are stationary. The control law (\ref{eq:PD}) is applied with the parameter $k = 1$. Fig. \ref{fig:double_stationary}a--\ref{fig:double_stationary}d illustrate that, with different initial positions, the agent will ultimately reach the Fermat-Weber point $\m{p}^* = [0,~0,~0]^\top$.

\textit{Simulation 2b - Beacons move with constant velocity:} 
In this simulation, the constant velocity of beacons are chosen as $\m{v}^* = \begin{bmatrix}
0.5,~0.3,~0.4 \end{bmatrix}^\top$ (m/s).
The initial position of the agent is $\m{p}(0)=[5,~5,~-3]^\top.$ Under control law (\ref{eq:PD adaptive}) with $k_1=1$ and $ k_2=1$, it can be seen from Fig. \ref{fig:double_constant}a--\ref{fig:double_constant}d that the agent eventually tracks the moving minimum, i.e., $\m{p}(t)\to {\m{p}}^*$ and $\m{v}(t) \to \m{v}^*$ as $t\to \infty$. 

\begin{figure*}[t]
\centering
\subfloat[]{\includegraphics[width=.7\linewidth]{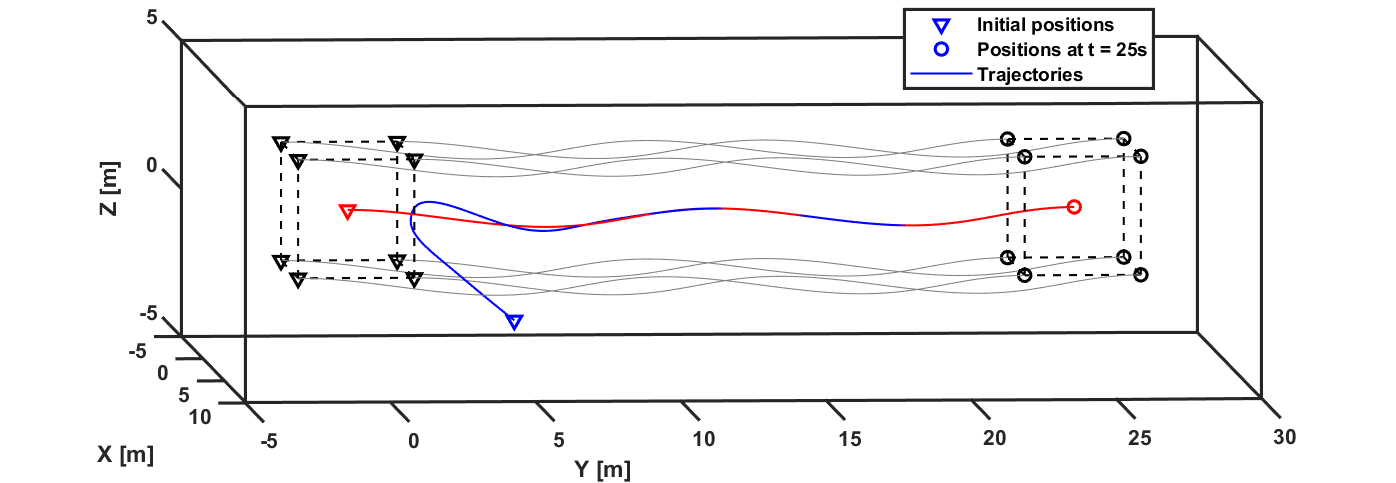}}
\\
\subfloat[]{\includegraphics[width=0.32\linewidth]{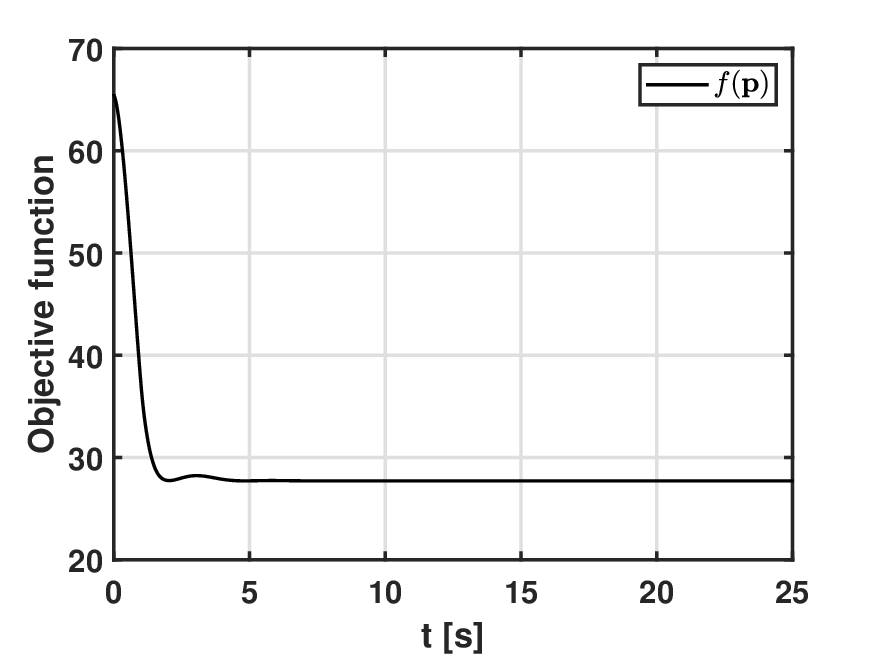}}
\hfill
\subfloat[]{\includegraphics[width=0.32\linewidth]{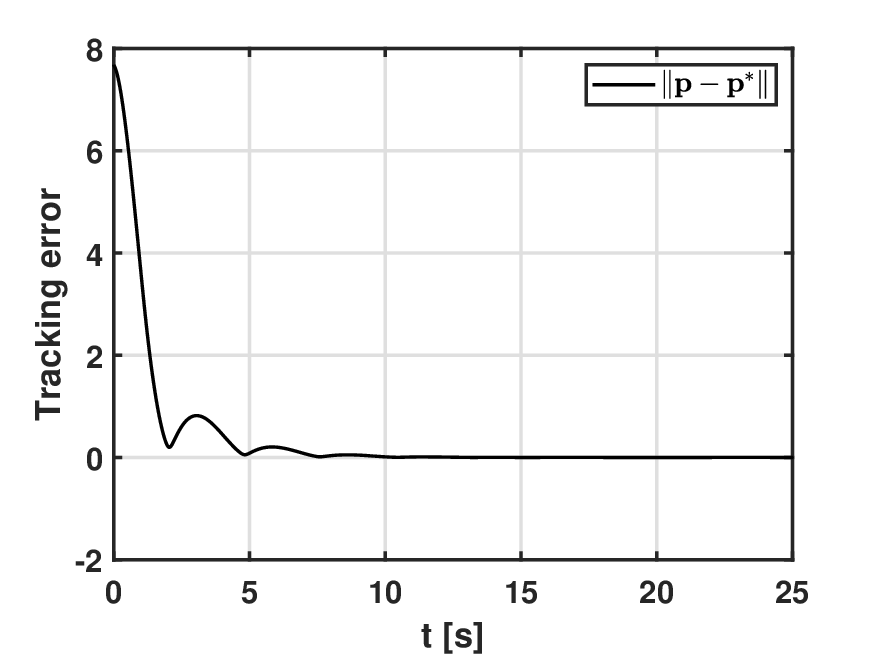}}
\hfill
\subfloat[]{\includegraphics[width=0.32\linewidth]{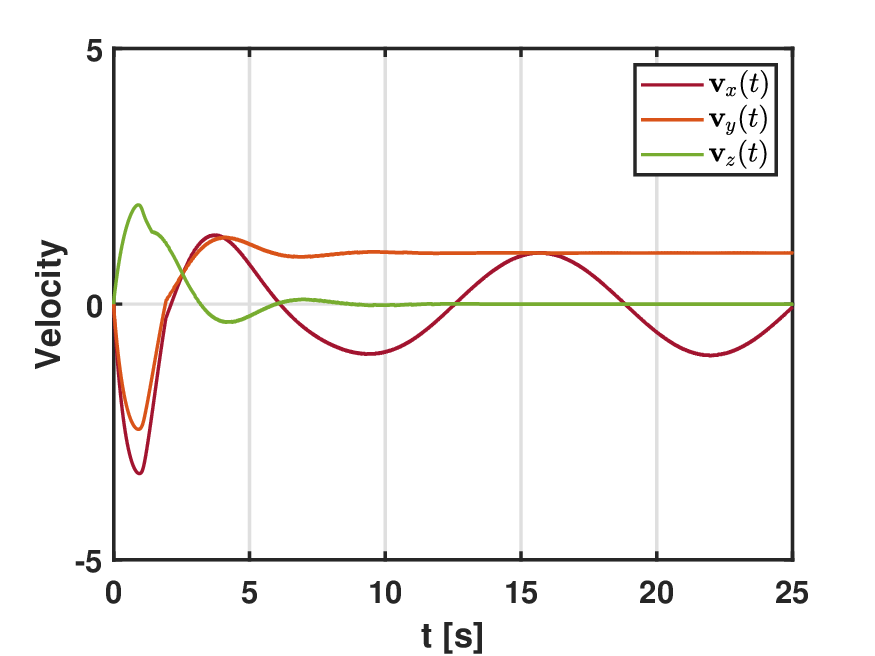}}
\caption{Simulation 2c: Double-integrator agent model and moving beacons with time-varying velocity. (a) The trajectories of the beacons, the agent, and the Fermat-Weber point are colored in black, blue, and red, respectively. Their positions at $t = 0$ and $t = 25$ sec are marked with $'\triangledown'$ and $'\circ'$, respectively; (b) The  function $f(\m{p})$ versus time; (c) The norm of tracking error $\|\bm{\delta}(t)\|$ versus time; (d) The velocities of the agent along the x, y, and z axes. }\label{fig:double_moving_time}
\end{figure*}

\textit{Simulation 2c - Beacons move with time-varying velocity:} 
In this simulation, the velocity of beacons are chosen as $\m{v}^* = [\text{sin}(\frac{t}{2}),~1,~0]^\top$ (m/s). With $\beta = 1$, simulation results are shown in Fig.~ \ref{fig:double_moving_time}a--\ref{fig:double_moving_time}d. It is clearly seen from Fig.~\ref{fig:double_moving_time}b--\ref{fig:double_moving_time}c that under control law (\ref{eq:moving 2nd order}), the tracking error $\bm{\delta}(t)$ will eventually converge to 0. Furthermore, Fig. \ref{fig:double_moving_time}d shows that  $\m{v}(t) \to \m{v}^*(t)$ as $t\to \infty$.

\section{Conclusion}\label{sec:conclusion}
This paper proposes novel bearing-only solutions for Fermat-Weber location problem. 
Building on the existing algorithm with a single-integrator agent model and stationary beacons, we introduce additional results that include global exponential stability, robustness against measurement noise, and a finite-time algorithm. 
Additionally, we consider the double-integrator agent model. For both types of dynamics, we study the algorithms with moving beacons, considering scenarios with both constant and time-varying velocities. Future work will focus on extending these results to handle nonlinear agent dynamics and negative weights in the objective function, i.e., attraction–repulsion problem.


\bibliographystyle{abbrv}        
\bibliography{autosam}           



\appendix



\end{document}